\documentclass[12pt,superscriptaddress,prc,nofootinbib]{revtex4}

\usepackage[british]{babel}
\usepackage[dvips,usenames]{color}              
\usepackage[final]{graphicx}            
\graphicspath{{figures/}{./}}

\usepackage{multirow}                   
\usepackage{amssymb}
\usepackage{amsmath}
\usepackage{xspace}                    
\usepackage{dcolumn}                    
\usepackage{bm}                        
\usepackage{bbm}

\usepackage{epstopdf}

\newcommand{\dis}{\displaystyle}

\newcommand{\hq}{\hspace{0.5em}}

\newcommand{\half}{\frac{1}{2}}
\newcommand{\e}{\mathrm{e}}
\newcommand{\ii}{\mathrm{i}}
\newcommand{\dd}{\mathrm{d}}

\newcommand{\deint}[2]{\dd^{#1}\! #2\;}

\newcommand{\ev}{\vec{e}}
\newcommand{\kv}{\vec{k}}
\newcommand{\pv}{\vec{p}}
\newcommand{\qv}{\vec{q}}

\newcommand{\MeV}{\ensuremath{\mathrm{MeV}}}
\newcommand{\GeV}{\ensuremath{\mathrm{GeV}}}
\newcommand{\fm}{\ensuremath{\mathrm{fm}}}
\newcommand{\EFTNoPion}{EFT(${\pi\hskip-0.55em /}$)\xspace}

\newcommand{\NXLO}[1]{N\ensuremath{{}^{#1}}LO\xspace}
\newcommand{\NtwoLO}{\NXLO{2}}

\newcommand{\wave}[3]{\ensuremath{{}^{#1}\mathrm{#2}_{#3}}\xspace}
\newcommand{\oneS}{\wave{1}{S}{0}}
\newcommand{\twoS}{\wave{2}{S}{\half}}
\newcommand{\threeS}{\wave{3}{S}{1}}
\newcommand{\fourS}{\wave{4}{S}{\frac{3}{2}}}
\newcommand{\twoPone}{\wave{2}{P}{\half}}
\newcommand{\twoPthree}{\wave{2}{P}{\frac{3}{2}}}
\newcommand{\fourPone}{\wave{4}{P}{\half}}
\newcommand{\fourPthree}{\wave{4}{P}{\frac{3}{2}}}

\newcommand{\LambdaNoPion}{\ensuremath{\Lambda_{\pi\hskip-0.4em /}}}

\newcommand{\ND}{N^\dagger}

\newcommand{\VS}{\vec{\sigma}}

\newcommand{\LRd}{\overset{\leftrightarrow}{\de}}

\newcommand{\daR}{g^{(^3 \! S_1-^1 \! P_1)}}
\newcommand{\dbR}{g^{(^1 \! S_0-^3 \! P_0)}_{(\Delta I=0)}}
\newcommand{\dcR}{g^{(^1 \! S_0-^3 \! P_0)}_{(\Delta I=1)}}
\newcommand{\ddR}{g^{(^1 \! S_0-^3 \! P_0)}_{(\Delta I=2)}}
\newcommand{\deR}{g^{(^3 \! S_1-^3 \! P_1)}}

\renewcommand{\Re}{\mathrm{Re}}

\newcommand{\de}{\partial}

 \newcommand{\calD}{\mathcal{D}}
\newcommand{\calE}{\mathcal{E}}
 \newcommand{\calK}{\mathcal{K}}
 \newcommand{\calM}{\mathcal{M}}
 \newcommand{\calO}{\mathcal{O}}
\newcommand{\calP}{\mathcal{P}} 
\newcommand{\calS}{\mathcal{S}} \newcommand{\calT}{\mathcal{T}}
\newcommand{\calZ}{\mathcal{Z}}

\newcommand{\nup}{\text{n}\!\! \uparrow}
\newcommand{\ndown}{\text{n}\!\! \downarrow}
\newcommand{\pup}{\text{p}\!\! \uparrow}
\newcommand{\pdown}{\text{p}\!\! \downarrow}

\newcommand{\gdiamond}{{\color{Green}\ensuremath{\rotatebox{45}{ \rule{1ex}{1ex} }}}\xspace}
\newcommand{\bdisc}{{\color{Blue}\ensuremath{\bullet}}\xspace}
\newcommand{\gtriangle}{{\color{Green}\ensuremath{\blacktriangle}}\xspace}
\newcommand{\rsquare}{{\color{Red}\ensuremath{ \rule{1ex}{1ex} }}\xspace}
\newcommand{\rcross}{{\color{Red}\ensuremath{+}}\xspace}
\newcommand{\rcirc}{{\color{Red}\ensuremath{\circ}}\xspace}
\newcommand{\rx}{{\color{Red}\ensuremath{\times}}\xspace}

\begin{document}

\title{Parity-violating neutron spin rotation in hydrogen and deuterium}

\author{Harald W.~Grie{\ss}hammer} \email{hgrie@gwu.edu} \affiliation{Institute
  for Nuclear Studies, Department of Physics, The George Washington
  University, Washington DC 20052, USA} \author{Matthias R.~Schindler}
\email{schindler@sc.edu} \affiliation{Institute for Nuclear Studies, Department
  of Physics, The George Washington University, Washington DC 20052, USA}
\affiliation{Department of Physics and Astronomy, University of South
  Carolina, Columbia, SC 29208, USA} \author{Roxanne P.~Springer}
\email{rps@phy.duke.edu} \affiliation{Department of Physics, Duke University,
  Durham, NC 27708, USA}   

\date{26th September 2011}

\begin{abstract}
  We calculate the (parity-violating) spin rotation angle of a polarized
neutron beam through hydrogen and deuterium targets, using pionless
effective field theory up to next-to-leading order. Our
result is part of a program to obtain the five leading independent
low-energy parameters that characterize hadronic parity-violation from
few-body observables in one systematic and consistent framework. The two
spin-rotation angles provide independent constraints on these
parameters. 
Our result for $np$ spin rotation is 
$\frac{1}{\rho}\;\frac{\dd \phi_\text{PV}^{np}}{\dd l} =  [4.5 \pm 0.5]\,  {\rm rad}\, \MeV^{-\half}
  \, \left(2\deR+\daR\right)  - [18.5 \pm 1.9]\,  {\rm rad} \, \MeV^{-\half} \, \left(\dbR - 2 \ddR\right)$,
while for $nd$ spin rotation we obtain
$\frac{1}{\rho} \frac{\dd \phi_\text{PV}^{nd}}{\dd l}
  = [8.0\pm0.8]\,  {\rm rad} \, \MeV^{-\half}\;\daR\;+\;[17.0\pm1.7]\,  {\rm rad} \, \MeV^{-\half}\;\deR+[2.3\pm0.5]\,  {\rm rad} \, \MeV^{-\half}\;\left(3\dbR -2\dcR\right)$, where the $g^{(X-Y)}$, in units of $\MeV^{-\frac{3}{2}}$, are the presently unknown parameters in the leading-order parity-violating Lagrangian. Using na\"ive dimensional analysis to estimate the
typical size of the couplings,
we expect the signal for standard target densities to be
$\left|\frac{\dd \phi_\text{PV}}{\dd l}\right|\approx \left[10^{-7}\cdots10^{-6}\right]\; \frac{\text{rad}}{\text{m}} $ for both hydrogen and deuterium targets. We find no indication that the $nd$ observable is enhanced compared to the $np$ one.  All results are properly renormalized. An estimate of the numerical and systematic uncertainties of our calculations indicates excellent convergence.  
An appendix contains the relevant partial-wave projectors of the three-nucleon system.
\end{abstract}

\maketitle

\section{Introduction}\label{Sec:Intro}
\setcounter{equation}{0}

Parity-violating nucleon-nucleon interactions cause the spin of
transversely polarized neutrons to undergo a rotation when traveling through
a target medium, even in the absence of magnetic fields. In this paper
we report the results of a calculation of neutron spin rotation
from neutron-proton and
neutron-deuteron forward scattering using pionless effective field
theory, \EFTNoPion. This work is part of an effort to provide one
consistent \EFTNoPion framework with reliable theoretical
uncertainties to parity-violating (PV) interactions in few-nucleon
systems. We hope that the results
presented here,
along with the results from Refs.~\cite{Phillips:2008hn,Schindler:2009wd}, will assist in the planning, analysis, and
interpretation of related PV experiments. 
 
The PV component of the force between nucleons stems from the weak interactions between the
standard model constituents of the nucleons. Compared to the parity-conserving (PC) part, it is typically suppressed by a factor of
$\sim 10^{-7}$ to $10^{-6}$; for reviews see e.g.~Refs.~\cite{Adelberger:1985ik,RamseyMusolf:2006dz}.
Parity-violating neutron spin rotation observables were first
discussed by Michel in 1964~\cite{Michel:1964zz} and studied further in
Refs.~\cite{Stodolsky:1974hm,Stodolsky:1980dg,Stodolsky:1981vn}. Recently, an
upper bound on the effect in Helium-4 was obtained at NIST~\cite{Snow:2011zz}. 

At present, the effects of hadronic parity violation on the nuclear level cannot be predicted from first principles. Early approaches to PV nucleon-nucleon ($NN$) interactions include the parameterization in terms of S-P wave transitions \cite{Danilov} and the more widely used phenomenological meson-exchange models, particularly the framework developed by Desplanques, Donoghue and Holstein (DDH) in Ref.~\cite{Desplanques:1979hn}. The DDH approach provides estimated ranges for PV meson-nucleon couplings based on a number of model assumptions. Using the ``best values,'' it has been applied to the study of neutron spin rotation on various targets~\cite{Dmitriev:1983mg,Avishai:1985mu,Schiavilla:2004wn,Schiavilla:2008ic}.  Calculations have also been
performed in the so-called hybrid formalism \cite{Liu:2006dm,Schiavilla:2008ic,Song:2010sz}, where phenomenological wave functions in the strong 
sector are combined with a PV effective field theory (EFT) treatment. We discuss this further in Sec.~\ref{Sec:Comp}.

In order to avoid model assumptions and to treat all interactions within a unified framework, we apply effective field theory techniques consistently to the neutron-hydrogen and neutron-deuterium systems. The identification of a small parameter allows a systematic expansion of our results and
a reliable estimate of the size of theoretical errors. In particular, since typical neutron energies in parity-violating spin-rotation experiments on the lightest nuclei are low enough that pion exchange cannot be resolved, we use the pionless EFT with only nucleons as dynamical degrees of freedom. This theory has proven highly
successful in the parity-conserving sector; for an overview see e.g.~Refs.~\cite{Beane:2000fx,Bedaque:2002mn,Platter:2009gz}. For early EFT descriptions of hadronic parity violation see Refs.~\cite{Kaplan:1992vj,Savage:1998rx,Kaplan:1998xi}. A comprehensive
formulation of PV effects in EFTs with and without pions was given in
Ref.~\cite{Zhu:2004vw}.

In the PV sector of \EFTNoPion, five independent operators appear at leading order (LO) in \EFTNoPion. They correspond to
the five transition amplitudes from Ref.~\cite{Danilov} expressed in
a field theory language.
The five accompanying parameters, or low-energy constants (LECs), encode the unresolved short-distance physics. At present, only experimental input can determine these
couplings without introducing additional model dependence.
Note, however, that for the pionful sector a first study using lattice simulations to determine the PV $\pi NN$ coupling has been performed \cite{Wasem:2011tp}.
Our final, next-to-leading order (NLO) results for $np$ and $nd$ spin rotation provide these processes in terms of  the PV LECs, along with estimates of the associated theoretical errors. Measurements of these observables can  determine two independent combinations of the PV low-energy constants.

This article is organized as follows: We first review the general formalism of
neutron spin rotation in Sec.~\ref{Sec:Rot} and present the necessary
PC and PV pieces of the Lagrangian in
Secs.~\ref{Sec:PC} and~\ref{Sec:PVLag}, respectively. The results for neutron
spin-rotation on the proton up to NLO, along with error estimates, are given in Sec.~\ref{Sec:npresult}.  The results for deuterium up to NLO are derived with a
detailed discussion of the expected theoretical uncertainties in
Sec.~\ref{Sec:ndSpinRot}.
In Sec.~\ref{Sec:Comp} we estimate numerical predictions and compare
with earlier work. Conclusions and outlook are given in Sec.~\ref{Sec:Outlook}.
Appendices contain the general construction principle for the
partial-wave projectors of the three-nucleon system and its results for the
$\text{S}$- and $\text{P}$-waves as well as details of the numerical calculations.\footnote{After completion of this article a paper appeared by Vanasse \cite{arXiv:1110.1039}, with an analysis of neutron spin rotation off the deuteron using pionless EFT to leading order.}

\section{Neutron spin rotation -- general formalism}\label{Sec:Rot}

In this section, we define the spin rotation angle, its relation to the
scattering amplitude, and the associated conventions we will use. Important
resources are Refs.\cite{Michel:1964zz,Fermi,Stodolsky:1974hm,Stodolsky:1980dg,Stodolsky:1981vn},
but note that conventions vary. 

A beam of very-low energy neutrons passing through
a medium picks up a phase factor from scattering in the target. This
phase factor is related to the index of refraction $n$ of the medium.
In the simplest case of very low-energy scattering with
plane waves describing the incoming state, the phase accumulated after
traversing a target of thickness $l$ is given by 
\begin{equation}
  \varphi=\Re(n-1)k_\text{lab}l,
\end{equation}
where $k_\text{lab}$ is the magnitude of the wave vector of the incoming
particle. The index of refraction is in turn related to the scattering length
$a$ by
\begin{equation}
  n-1=-\frac{2\pi \rho\, a }{k_\text{lab}^2},
\end{equation}
where $\rho$ is the density of scattering centers in the target.
In our convention, the scattering length and scattering amplitude at zero energy are related by
\begin{equation}
  \calM=-\frac{2\pi}{\mu}a,
\end{equation}
with $\mu$ the reduced mass of the beam-target system. The phase $\varphi$ can therefore be written as
\begin{equation}\label{eq:phase}
  \varphi=\rho l \frac{\mu}{k_\text{lab}}\Re(\calM)\;.
\end{equation}

For a beam chosen in the $+z$ direction,
a perpendicular polarization in the $+x$ direction can be written as
\begin{equation}
  \lvert x_+ \rangle = \frac{1}{\sqrt{2}}\left( \lvert + \rangle + \lvert -
    \rangle \right), 
\end{equation}
where the states $\lvert \pm \rangle$ represent states with positive/negative helicity along $+\hat z$. 
When traversing a
medium, each helicity state evolves with a phase factor: 
\begin{equation}
  \frac{1}{\sqrt{2}} \left(e^{-\ii\phi_+}\lvert + \rangle + e^{-\ii\phi_-}
    \lvert - \rangle \right) = \frac{1}{\sqrt{2}} e^{-\ii\phi_+} \left(\lvert + \rangle + e^{\ii(\phi_+ - \phi_-)}
    \lvert - \rangle \right). 
\end{equation}
As long as $\phi_+ = \phi_-$, which is the case for parity-conserving
interactions, the polarization of the beam is unchanged; the state simply
picks up an overall phase factor. In the case of parity violation, however, $\phi_+ \ne \phi_-$ and the neutron
spin is rotated by an amount
\begin{equation}
  \phi_\text{PV}=\phi_+ - \phi_-.
\end{equation}
A positive value of the spin rotation angle $\phi_\text{PV}$ corresponds to a rotation about the
neutron momentum in the sense of a right-handed screw. 
Using Eq.~\eqref{eq:phase}, the
spin rotation angle per unit length $l$ is
\begin{equation}
\label{eq:spinrotation}
\frac{1}{\rho}\;\frac{\dd \phi_\text{PV}}{\dd l} = \frac{\mu}{k_\text{lab}}\, \Re\left(\calM_+ - \calM_- \right),
\end{equation}
where $\calM_\pm$ is the scattering amplitude for $\pm$-helicity neutrons including the statistical mixture of available target spins.  For further details see Refs.~\cite{Stodolsky:1974hm,Stodolsky:1981vn}.

\section{Parity-conserving Lagrangians and amplitudes}\label{Sec:PC}
\setcounter{equation}{0}

\subsection{Two-Nucleon Sector}

A description of \EFTNoPion and its power counting can be found in
Refs.~\cite{Beane:2000fx,Bedaque:2002mn,Platter:2009gz}, for example.  Pionless EFT is
applicable to energies $E<m_\pi^2/M$, where $m_\pi$ and $M$ are the pion and
nucleon mass, respectively. The short-distance details of the interactions are
encoded in the low-energy couplings (LECs) of $NN$ contact interactions.
Operators and observables are organized in terms of the small
dimensionless expansion parameter
$Q=\frac{p_\text{typ}}{\LambdaNoPion}$, where $p_\text{typ}$ is the typical
external momentum or momentum transfer and $\LambdaNoPion\sim m_\pi$ is the
breakdown scale of the theory (the scale at which pion exchange can be
resolved). The expansion parameter is typically 1/5 to 1/3. The following leading-order (LO) and next-to-leading order (NLO)
calculations only require S-wave interactions in the parity-conserving $NN$
sector; higher partial waves are suppressed according to the power counting.
It is convenient (in particular for three-body calculations) to introduce
spin-triplet and spin-singlet dibaryon fields $d_t$ and $d_s$ with the quantum
numbers of the corresponding two-nucleon \wave{}{S}{}-wave states
\cite{Kaplan:1996nv,Bedaque:1999vb,Beane:2000fi}. The dibaryon (auxiliary) field $d_t$ also serves as
the deuteron interpolating field since both have identical quantum
numbers.  The relevant terms of the Lagrangian up to NLO  are
\begin{align}
  \label{eq:PCLag}
  \mathcal{L}_{PC} =& \ND\left(\ii \partial_0 +
    \frac{\vec{\partial}^2}{2M}\right)N -y\left[ d_t^{i\dagger} (N^T P^i_t N)
    +\mathrm{H.c.}\right]-y\left[ d_s^{A \dagger} (N^T P^A_s N)
    +\mathrm{H.c.}\right] \notag\\
  & +d_t^{i\dagger}\left[\Delta_t
    -c_{0t}\left(\ii\partial_0+\frac{\vec{\partial}^2}{4M}+\frac{\gamma_t^2}{M}\right)
  \right] d_t^i +d_s^{A\dagger}\left[\Delta_s -c_{0s}\left(\ii\partial_0+
      \frac{\vec{\partial}^2}{4M}+\frac{\gamma_s^2}{M}\right)\right] d_s^A
  \notag \\   & + \ldots,
\end{align}
using the conventions of Ref.~\cite{Griesshammer:2004pe}. 
$N=\binom{p}{n}$ is the isospin doublet of nucleon Weyl spinors $p$ (proton)
and $n$ (neutron).  With $\sigma_i$ ($\tau_A$) the Pauli matrices in spin
(isospin) space, the matrices $P_t^i=\frac{1}{\sqrt{8}}\tau_2\sigma_2\sigma_i$
and $P_s^A=\frac{1}{\sqrt{8}}\tau_2\tau_A\sigma_2$ project two-nucleon states
onto the \wave{3}{S}{1} and \wave{1}{S}{0} partial waves~\cite{Kaplan:1998sz}.

The parameters of the Lagrangian are fixed using Z-parameterization
\cite{Phillips:1999hh,Griesshammer:2004pe}. Choosing
\begin{equation}
  \label{eq:y}
y^2=\frac{4\pi}{M}\;\;,
\end{equation} 
the LO parameters $\Delta_{s/t}$ are determined from the poles of
the $NN$ \wave{}{S}{}-wave scattering amplitudes at $\ii\gamma_{s/t}$. In the
\threeS-wave, this reproduces the experimental binding energy of the deuteron
$B_d=-\gamma_t^2/M$. The leading-order
dibaryon propagators are then given by
\begin{equation}
  \label{eq:DibProp}
  D^\text{LO}_{s/t}(q_0,\qv) =
  \frac{1}{\gamma_{s/t}-\sqrt{\frac{\qv^2}{4}-Mq_0-\ii\epsilon}}\;\;. 
\end{equation}
At NLO, only the additional parameters $c_{0s/t}$ enter. In
$Z$-parameterization, they are chosen such that the residues of the poles in the dibaryon propagators do not receive any corrections beyond NLO; see again
Ref.~\cite{Griesshammer:2004pe} for details:
\begin{equation}
\label{eq:c0}
  c_{0s/t}=-\frac{M}{2\gamma_{s/t}}\;(Z_{s/t}-1)\;\;, 
\end{equation}
where $Z_{s/t}=1/(1-\gamma_{s/t}\rho_{s/t})$ is related to the effective range
$\rho_{s/t}$. Up to NLO, the dibaryon propagators in Eq.~\eqref{eq:DibProp}
are modified by an insertion of the effective-range term $c_{0s/t}$ of the
Lagrangian of Eq.~(\ref{eq:PCLag}):
\begin{equation}
  \label{eq:calDNLO}
  D^\text{LO+NLO}_{s/t}(q_0,\qv)= D^\text{LO}_{s/t}(q_0,\qv)+
  D^\text{NLO}_{s/t}(q_0,\qv)\;\;,
\end{equation}
with the NLO correction (Z-parameterization variant in the second line)
\begin{align}
  \label{eq:calDNLOcorr}
  D^\text{NLO}_{s/t}(q_0,\qv)&= \ii D^\text{LO}_{s/t}(q_0,\qv)\;
  (-c_{0s/t})\left(q_0-\frac{q^2}{4M}+\frac{\gamma_{s/t}^2}{M}\right)
  \; \ii D^\text{LO}_{s/t}(q_0,\qv)\\
  &=\frac{\gamma_{s/t}+\sqrt{\frac{\qv^2}{4}-Mq_0-\ii\epsilon}}
  {\gamma_{s/t}-\sqrt{\frac{\qv^2}{4}-Mq_0-\ii\epsilon}}\;
\frac{Z_{s/t}-1}{2\gamma_{s/t}}\nonumber\;\;.
\end{align}

Calculations with a deuteron as an external state require wave function
renormalization, $\calZ_{t}$.  In Z-parameterization, the LO expression is
\begin{equation}
\label{eq:calZLO}
  \calZ_{t}^\text{LO}=\frac{2\gamma_{t}}{M}\;,
\end{equation}
and up to NLO
\begin{equation}
  \label{eq:calZNLO}
  \calZ_{t}^\text{LO+NLO}=\frac{2\gamma_{t}}{M}\;Z_t = \frac{2\gamma_{t}}{M}\;\frac{1}{(1-\gamma_{t}\rho_{t})}\;.
\end{equation}
With $Z_t=1.6908$, the NLO correction results in a $70\%$-shift from the LO
value. While this contribution is much larger than expected from na{\"i}ve
power counting, there are no further corrections to $\calZ_{t}$ at higher
orders. This has the important advantage that the correct asymptotic
normalization of the deuteron wave function at large distances $r$ is exactly
reproduced at NLO, with no corrections at higher orders:
\begin{equation}
  \Psi_\text{deuteron}(r\to\infty)=\sqrt{\frac{\gamma_tZ_t}{2\pi}}\;
  \frac{\e^{-\gamma_tr}}{r} \;.
\end{equation}
Taking into account the unusually large NLO term therefore significantly
increases overall convergence of the expansion of \EFTNoPion, as demonstrated
e.g.~in Refs.~\cite{Phillips:1999hh,Griesshammer:2004pe}.

\subsection{Three-Nucleon Sector}
\label{sec:3NPC}

The consistency requirement of including a three-nucleon interaction (3NI) in
the \twoS channel even at LO is discussed in reviews; see
e.g.~Refs.~\cite{Beane:2000fx,Bedaque:2002mn,Platter:2009gz}.  The corresponding
Lagrangian is given by
\begin{equation}
  \label{eq:3NLag}
  \mathcal{L}_{3N}=\frac{y^2M\,H_0(\Lambda)}{3\Lambda^2}
  \left[d^i_t(\sigma_iN)-d_s^A(\tau_AN)\right]^\dagger
  \left[d^i_t(\sigma_iN)-d_s^A(\tau_AN)\right],
\end{equation}
where $H_0(\Lambda)$ denotes the three-nucleon coupling, which depends on the
UV regulator $\Lambda$. This is the only parameter of $Nd$
scattering up to NLO  not determined from $NN$ experiments.  The 3NI
strength $H_0(\Lambda)$ can be chosen to reproduce the triton binding energy,
or the \twoS-scattering length.  Choosing different low-energy data to fix it
provides one method to estimate the theoretical uncertainties in
Sec.~\ref{sec:ndresult}.

\begin{figure}[!htbp]
  \begin{center}
    \includegraphics*{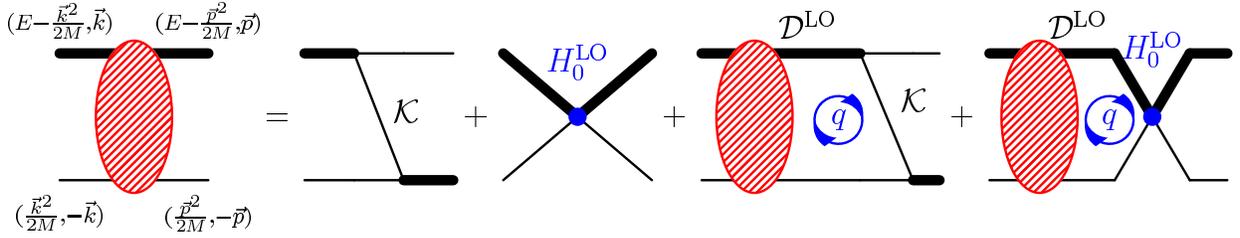}
    \caption{Nucleon-deuteron scattering at LO. Shaded ellipse: LO amplitude
      $t^\text{LO}$; thick line: LO dibaryon propagator $\calD^\text{LO}$ of Eq.~\eqref{eq:calD}; thin line
      ($\calK$): propagator of the exchanged nucleon; disc: PC 3NI
      ($H_0^\text{LO}$).}
    \label{fig:3NAmpt}
  \end{center}
\end{figure}

The parity-conserving $nd$ scattering amplitude
is found by solving a Faddeev equation, see e.g.~Ref.~\cite{Griesshammer:2004pe}.
Its pictorial representation in Fig.~\ref{fig:3NAmpt} specifies the
center-of-mass kinematics: The total non-relativistic energy is $E$, and the
momentum of the incoming (outgoing) deuteron is $\kv$ ($\pv$). The on-shell
point is at $E=\frac{3{\kv}^2}{4M}-\frac{\gamma_t^2}{M}+\ii\epsilon$ and $p=k$.

In the dibaryon framework, the three-nucleon system receives contributions
from $Nd_s$ and $Nd_t$ configurations, which are conveniently taken into
account by decomposing all operators and amplitudes into the so-called
cluster-decomposition basis (see Ref.~\cite[App.~A.1]{Griesshammer:2004pe} and
our App.~\ref{Sec:Projectors}),
\begin{equation}
  \label{eq:clusterdecomp}
  \calO = N^\dagger_{b\beta} \;\left(d^\dagger_{t,j},\;d^\dagger_{s,B}\right)
  \begin{pmatrix}\calO(Nd_t\to Nd_t)^j_i&\calO(Nd_s\to Nd_t)^j_A\\
    \calO(Nd_t\to Nd_s)^B_i&\calO(Nd_s\to Nd_s)^B_A
  \end{pmatrix}^{b\beta}_{\hq\hq a\alpha} \;\binom{d_t^i}{d_s^A}\;N^{a\alpha},
\end{equation}
where the spinor and isospinor indices $(\alpha,\beta)$ and $(a,b)$,
respectively, are often suppressed in the following.  The vector indices
$i,j=(1,2,3)$ label the spin Pauli matrices while $A,B=(1,2,3)$ label the
isospin Pauli matrices. As an example, the LO dibaryon propagator in the 
cluster-decomposition basis is defined by a diagonal matrix in terms of the
dibaryon propagators Eq.~\eqref{eq:DibProp}:
\begin{equation}
  \label{eq:calD}
  \calD^\text{LO}(q_0,\qv)= \begin{pmatrix} D^\text{LO}_t(q_0,\qv) & 0 \\ 0 &
    D^\text{LO}_s(q_0,\qv)  \end{pmatrix}\;\;. 
\end{equation}
An analogous expression holds for the NLO correction of
Eq.~\eqref{eq:calDNLOcorr}.

The spin-quartet channel only receives contributions from $Nd_t \to Nd_t$, and
the corresponding amplitude $t_q$ is the solution to an integral equation that only
involves the $(11)$-element of the cluster matrix (the argument ($E;p_\text{in},p_\text{out})$
applies to each entry in the matrix):
\begin{align}
  \label{eq:quartetfaddeev}
  \begin{pmatrix}t_q^{(L)}&0\\0&0\end{pmatrix}
  (E;k,p)=&-4\pi \calK^{(L)}(E;k,p)\begin{pmatrix} 1 &0 \\ 0 & 0
  \end{pmatrix}\nonumber\\
  &+\frac{2}{\pi}\int_0^\Lambda \dd q\, q^2
  \calK^{(L)}(E;q,p)\begin{pmatrix} 1 &0 \\ 0 & 0 \end{pmatrix}
  \calD^\text{LO}(E;q)\;\begin{pmatrix}t_q^{(L)}&0\\0&0\end{pmatrix}(E;k,q),  
\end{align}
with $\Lambda$ the UV regulator. The projection of the exchange-nucleon
propagator onto a specific orbital angular momentum $L$ is
\begin{equation}\label{eq:NucProp}
  \calK^{(L)}(E;q,p)=\frac{1}{2} \int_{-1}^1 \dd\cos\theta
  \frac{P_L(\cos\theta)}{p^2+q^2-ME+pq\cos\theta}  
  = \frac{(-1)^L}{pq}Q_L\left( \frac{p^2+q^2-ME}{pq} \right),
\end{equation}
with $P_L(z)$ and $Q_L(z)$ the $L$th Legendre polynomials of the first and
second kind with complex argument, respectively \cite{Gradshteyn}, and
$\theta=\angle(\pv;\qv)$.

In the spin-doublet channel, amplitudes with different cluster-decompositions mix,
so that with $t_{d,xy}^{(L)}$ denoting the amplitude for the process $Nd_x\to
Nd_y$ and $x,y\in\{s,t\}$:
\begin{align}
\label{eq:doubletfaddeev}
  \begin{pmatrix} t_{d,tt}^{(L)} & t_{d,st}^{(L)}\\ t_{d,ts}^{(L)} &
    t_{d,ss}^{(L)} \end{pmatrix}&(E;k,p) =
  2\pi \left[\calK^{(L)}(E;k,p) \begin{pmatrix} 1 &-3 \\ -3 & 1 \end{pmatrix}
    +\delta_0^L \frac{2H_0(\Lambda)}{\Lambda^2} \begin{pmatrix} 1 & -1 \\ -1 &
      1 \end{pmatrix} \right] \notag\\ 
  & -\frac{1}{\pi} \int_0^\Lambda \dd q \, q^2 \left[\calK^{(L)}(E;q,p)    \begin{pmatrix} 1 &-3 \\ -3 & 1 \end{pmatrix}
    +\delta_0^L \frac{2H_0(\Lambda)}{\Lambda^2} \begin{pmatrix} 1 & -1 \\ -1 &
      1 \end{pmatrix} \right] \notag\\ 
  & \quad\quad\quad\quad\quad\quad \times\calD^\text{LO}(E;q)
  \begin{pmatrix} t_{d,tt}^{(L)} & t_{d,st}^{(L)}\\ t_{d,ts}^{(L)} &
    t_{d,ss}^{(L)} \end{pmatrix}(E;k,q)\;\;.
\end{align}
Since there is no partial-wave mixing in the PC sector even at NLO, the quartet- and
doublet scattering amplitudes can be combined into one cluster matrix
\begin{equation}
t^\text{LO}[{}^{2S+1}L_{J};k,q]= \left\{
  \begin{array}{ll}\begin{pmatrix} t_q^{(L)} &0 \\0 &0 \end{pmatrix}(E;k,q)& \mbox{ for the
      spin-quartet, $S=\frac{3}{2}$}\\[5ex]
    \begin{pmatrix}  t_{d,tt}^{(L)} & t_{d,st}^{(L)}\\ t_{d,ts}^{(L)} &
    t_{d,ss}^{(L)} \end{pmatrix}(E;k,q)& \mbox{ for the  spin-doublet, $S=\frac{1}{2}$}
  \end{array}\right. \;\;,
\end{equation}
where $S$ is the spin, $L$ the orbital angular momentum and $J$ the total
angular momentum of the $^{2S+1}L_{J}$ partial wave considered.

\begin{figure}[!htbp]
  \begin{center}
    \includegraphics*{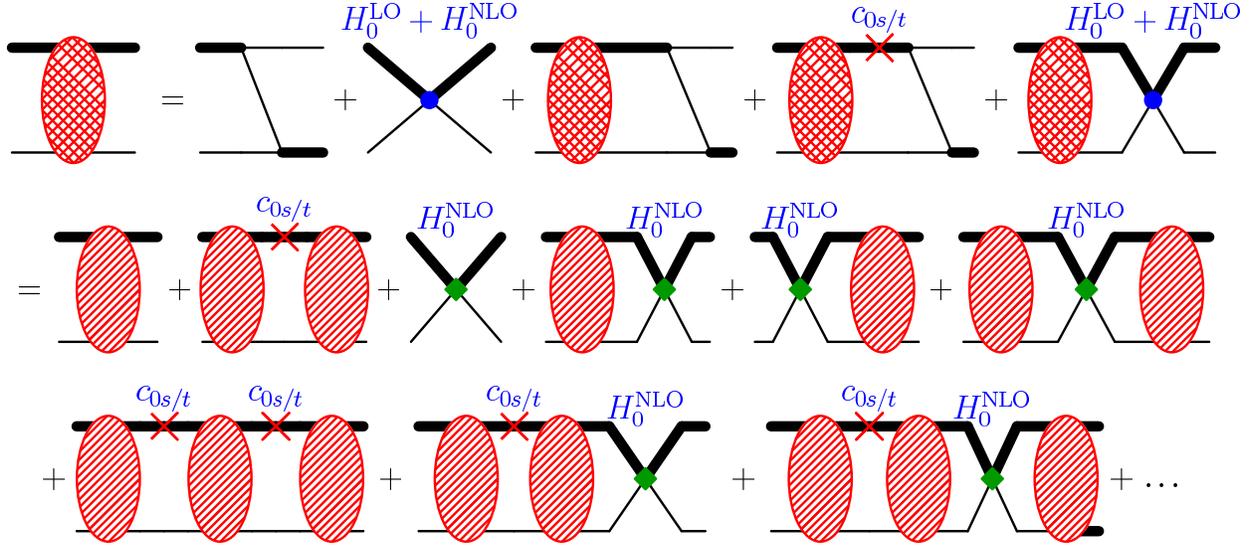}
    \caption{Top row: Nucleon-deuteron scattering in the partially-resummed
      formalism at NLO. Middle row: all LO and NLO
      contributions. Bottom row: some of the diagrams which are strictly 
      higher order than NLO but resummed for convenience.  Cross-hatched ellipse: $t^\text{LO+NLO}$ as defined in the text; cross: insertion of the effective-range
      correction $c_{0s/t}$ to the dibaryon propagator; diamond: PC 3NI $H_0^\text{NLO}$.}
    \label{fig:3NscatteringNLO}
  \end{center}
\end{figure}

\begin{figure}[!htb]
   \begin{center}
     \includegraphics*[width=0.48\linewidth]{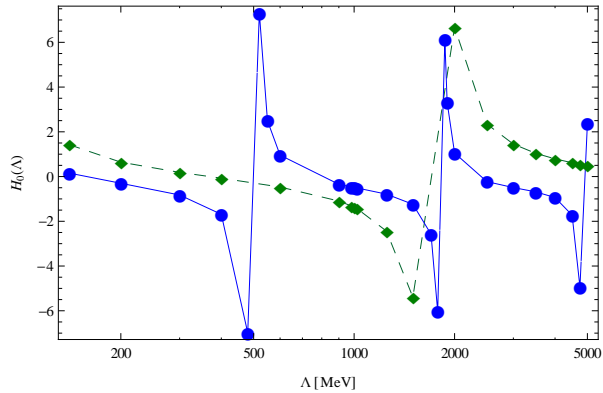}
     \caption{Cutoff dependence of the \twoS-wave PC 3NI $H_0(\Lambda)$,
determined to reproduce the \twoS scattering length. \protect\gdiamond
(dashed
lines): LO; \bdisc (solid lines): LO+NLO in the partially-resummed
formulation. The linear extrapolations are only meant to guide the eye.}
     \label{fig:H0dependence}
   \end{center}
\end{figure}

For the NLO PC amplitudes, we use the so-called ``partially-resummed'' formalism, in which the kernel and inhomogeneous part of the integral
equations are expanded to NLO and then iterated, see
Fig.~\ref{fig:3NscatteringNLO} and Ref.~\cite{Bedaque:2002yg}.
This modifies the dibaryon propagators in
Eqs.~(\ref{eq:quartetfaddeev}/\ref{eq:doubletfaddeev})
by an insertion of the effective-range contribution, i.e.~by replacing
$\calD^\text{LO}(E;q)$ with $\calD^\text{LO+NLO}(E;q)$, see
Eq.~\eqref{eq:calDNLO}. No new PC 3NI enters at NLO, but in order to reproduce the three-nucleon
observable, $H_0$ has to be adjusted at NLO. This is most
conveniently achieved by dividing $H_0$ into a LO piece $H_0^\text{LO}$
and a term at NLO, $H_0^\text{NLO}$.
In this approach, in addition to \emph{all} LO and NLO contributions, some higher-order contributions are also included in the amplitude referred to as $t^\text{LO+NLO}$. This does not increase the accuracy of the calculations,
which is still set by the order to which the kernel is expanded.
Figure~\ref{fig:H0dependence} shows the
cutoff dependence of the 3NI $H_0$ at LO and LO+NLO in the partially-resummed formulation.
While
$H_0$ varies considerably, observables are cutoff-independent, see
e.g.~\cite{Griesshammer:2004pe}. This point will be of importance in the
discussion of renormalization of the PV amplitudes in
Secs.~\ref{sec:ndNLO},
\ref{Sec:CutoffInd} and App.~\ref{app:insertions}.

The following parameters are used \cite{Griesshammer:2004pe}: $\hbar c=197.327\;\MeV\;\fm$, the isospin-averaged nucleon mass $M=938.918\;\MeV$; $\gamma_t=45.7025\;\MeV$,
$\gamma_s=-7.890\;\MeV$, $Z_t=1.6908$, and $Z_s=0.9015$ are the effective-range
parameters of the $NN$ system; and the $nd$ \twoS scattering length
$a_3=0.65\;\fm$~\cite{doublet_sca} or triton binding energy $B_3=8.48\;\MeV$ determines the 3NI $H_0(\Lambda)$.

\section{Parity-violating Lagrangian}\label{Sec:PVLag}
\setcounter{equation}{0}

The leading-order PV Lagrangian relevant to our calculation is given in terms
of five S$-$P wave transitions~\cite{Schindler:2009wd}, 
\begin{align}
  \label{eq:PVLag}
  \mathcal{L}_{PV} =& - \left[ \daR d_t^{i\dagger} \left(N^T
      \sigma_2 \tau_2\,\ii\LRd_i N\right) \right. \notag\\
  &\quad\quad +\dbR d_s^{A\dagger}
  \left(N^T\sigma_2 \ \VS \cdot \tau_2 \tau_A \,\ii\LRd  N\right) \notag\\
  &\quad\quad +\dcR \ \epsilon^{3AB} \, d_s^{A\dagger}
  \left(N^T \sigma_2  \ \VS\cdot \tau_2 \tau^B \LRd N\right) \notag\\
  &\quad\quad +\ddR \ \mathcal{I}^{AB} \, d_s^{A\dagger}
  \left(N^T \sigma_2 \ \VS\cdot \tau_2 \tau^B \,\ii \LRd N\right) \notag\\
  &\quad\quad \left. +\deR \ \epsilon^{ijk} \, d_t^{i\dagger} \left(N^T \sigma_2
      \sigma^k \tau_2 \tau_3 \LRd{}^{j} N\right) \right] +\mathrm{H.c.}
  +\ldots,
\end{align}
where $a\,\calO \LRd b= a\,\calO \overset{\rightarrow}{\partial} b -
(\overset{\rightarrow}{\partial}a)\calO b$, $\calO$ is a spin-isospin
operator, and $\mathcal{I}=\text{diag}(1,1,-2)$ is a diagonal matrix in
isospin space. For equivalent  Lagrangians in different bases see Refs.~\cite{Girlanda:2008ts,Phillips:2008hn}.

We note as an aside that the relation between these PV dibaryon couplings
$g^{(X-Y)}$ and the non-dibaryon couplings $\mathcal{C}^{(X-Y)}$ of
Ref.~\cite{Phillips:2008hn} differs from that given in
Ref.~\cite{Schindler:2009wd} due to different conventions used in the PC
sector. The general expression remains
\begin{equation}
\label{dict}
g^{(X-Y)} = \sqrt{8} \, \frac{\Delta_{s/t}}{y}\,\mathcal{C}^{(X-Y)}\;,
\end{equation}
with $\Delta_s$ for $X=\oneS$ and $\Delta_t$ for $X=\threeS$, but the
values for $\Delta_{s/t}$ and $y$ in the Z-parameterization used here
differ from the conventions of Ref.~\cite{Schindler:2009wd}.

Higher-order contributions to the Lagrangian of Eq.~\eqref{eq:PVLag} are suppressed by additional powers of $Q$. Corrections to S$-$P wave operators are expected to be suppressed by $Q^2$ since they contain the same spin-space structure as the terms in Eq.~\eqref{eq:PVLag}, but with two additional derivatives. The effects of different partial waves, such as P$-$D wave mixing, are suppressed even further.

As in the parity-conserving case, a simplistic application of the
power counting suggests that parity-violating 3NIs first start to appear at
\NXLO{2}.  Unlike the PC case, this simplistic power counting is valid for PV $nd$ scattering; parity-violating corrections to S$-$P wave
transitions from PV 3NIs do not contribute at LO or NLO
\cite{Griesshammer:2010nd}. The Lagrangian of Eq.~(\ref{eq:PVLag}) is
therefore sufficient to determine PV $nd$ scattering up to and including
NLO.

\section{Neutron-proton spin rotation}\label{Sec:npresult}
\setcounter{equation}{0}

\begin{figure}[!htb]
  \begin{center}
    \includegraphics*{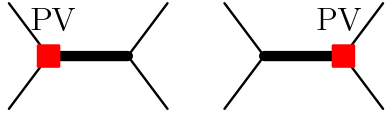}
    \caption{LO PV two-nucleon diagrams. Square: parity-violating
      two-nucleon vertices.}
    \label{fig:PV2N}
  \end{center}
\end{figure}

In the dibaryon formalism, the only diagrams contributing to $np$ spin rotation at LO
are the tree-level diagrams shown in Fig.~\ref{fig:PV2N}.
The corresponding non-zero amplitudes
are obtained by using the LO dibaryon propagator of
Eq.~(\ref{eq:DibProp}):
\begin{equation}\label{eq:npamplitudes}
\begin{split}
\ii {\cal M}[\nup \pup \rightarrow \nup \pup]  = &  \ii 8 \sqrt{2} k y \ \frac{\deR}
{\gamma_t+\ii k}\; ,\\
\ii {\cal M}[\ndown \pdown \rightarrow \ndown \pdown] = &
- \ii 8 \sqrt{2} k y \ \frac{\deR}{\gamma_t+\ii k} \; ,\\
\ii {\cal M}[\nup \pdown \rightarrow  \nup \pdown ] =&
\ii 4 \sqrt{2} k y \ \left( \frac{\daR}{\gamma_t+\ii k} +
 \frac{\dbR-2\ddR}{\gamma_s+\ii k}\right) \; ,
  \\  
\ii {\cal M}[\ndown \pup \rightarrow \ndown \pup] =& -
\ii 4 \sqrt{2} k y \ \left( \frac{\daR}{\gamma_t+\ii k} +
 \frac{\dbR-2\ddR}{\gamma_s+\ii k}\right)\;,
\end{split}
\end{equation}
with $k$ the magnitude of the
center-of-mass momentum.

Applying Eq.~(\ref{eq:spinrotation}) with $\mu=M/2$  for the reduced mass and $k_\text{lab}=2k$, the
spin rotation angle for polarized neutrons on a hydrogen target is given by
\begin{equation}
\label{eq:npspinrot}
\frac{1}{\rho}\;\frac{\dd \phi_\text{PV}}{\dd l} = \frac{M}{4k}
\sum_{m_p=\pm \frac{1}{2}}\Re\left[\calM_+(m_p) - \calM_-(m_p)
\right], 
\end{equation}
where $m_p=\pm\half$ is the (initial and final) proton polarization.
Since the (thermal) external nucleon
momentum $k$ appears only in the dibaryon propagators, it is negligible compared
to the parameters $\gamma_{s/t}$.  With Eq.~(\ref{eq:npamplitudes}), the PV rotation angle for a hydrogen target at
LO is  given by
\begin{align}
  \label{eq:npspinrotdecomposed}
  \frac{1}{\rho}\;\left.\frac{\dd \phi_\text{PV}^{np}}{\dd l}\right|_{\text{LO}}
  = & 2 \sqrt{2 \pi M} \left( \frac{2\deR+\daR}{\gamma_t} 
+ \frac{ \dbR-2 \ddR}{\gamma_s} \right) \\
  = & \left[ 3.4 \, \left(2\deR+\daR \right) -19.5 \, \left(\dbR -2 \ddR \right) \right] {\rm rad \ MeV^{-\half}} , \nonumber
\end{align}  
where the $g^{(X-Y)}$ carry units of MeV$^{-3/2}$.

\begin{figure}[!htb]
  \begin{center}
    \includegraphics*{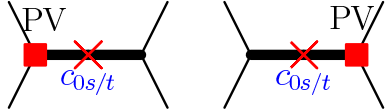}
    \caption{NLO PV two-nucleon diagrams.}
    \label{fig:PV2NNLO}
  \end{center}
\end{figure}
Since no new PV operators enter at NLO, as discussed in Sec.~\ref{Sec:PVLag},
the scattering amplitudes at NLO are given by the diagrams in
Fig.~\ref{fig:PV2NNLO}. Using the dibaryon propagator of
Eq.~(\ref{eq:calDNLO}), the LO+NLO result is obtained by multiplying
Eq.~\eqref{eq:npamplitudes} by $\frac{Z_{s/t}+1}{2}$ for each term
$\frac{1}{\gamma_{s/t}+\ii k}$ if
the neutron momentum is neglected compared to $\gamma_{s/t}$.  This correction
is 35\% for the P$-\threeS$ waves and 5\% for the P$-\oneS$ waves, in
agreement with the na\"ive power counting estimate of the \EFTNoPion expansion.  Our expression up to NLO becomes 
\begin{equation}
\label{eq:npspinrotresultNLO}
\begin{split}
\left.\frac{1}{\rho}\;\frac{\dd \phi_\text{PV}^{np}}{\dd l}\right|_{\text{LO+NLO}}
  =  & \bigg( [4.5 \pm 0.5] \, \left(2\deR + \daR \right) \\
& - [18.5 \pm 1.9] \, \left(\dbR -2 \ddR\right) \bigg) {\rm rad \ MeV^{-\half}}\;,
\end{split}
\end{equation}
where we have conservatively assigned errors of ${\cal O}(Q^2)\sim 0.1$.


\section{Neutron-deuteron spin rotation}\label{Sec:ndSpinRot}
\setcounter{equation}{0}

\subsection{$nd$ partial-wave amplitudes at leading order}
\label{sec:ndLO}

\begin{figure}[!htb]
  \begin{center}
    \includegraphics*{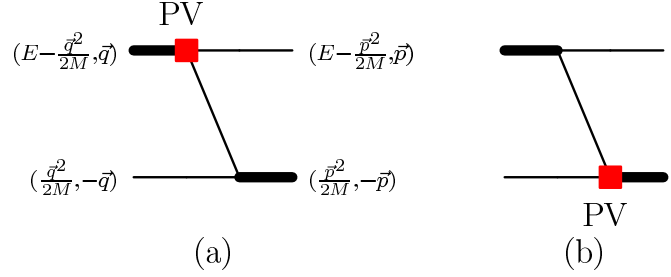}
    \caption{LO tree-level PV diagrams and off-shell kinematics for
      convolution in Figs.~\ref{fig:oneloop} and \ref{fig:twoloop}.}
    \label{fig:tree}
  \end{center}
\end{figure}

\begin{figure}[!htb]
  \begin{center}
    \includegraphics*{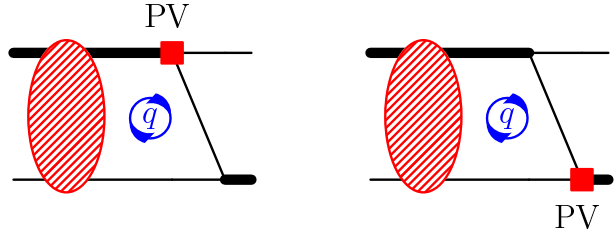}
    \caption{LO ``one-loop'' PV diagrams. ``Time-reversed'' contributions 
    (diagrams obtained by reading above ones from right to left)
    not displayed.}
    \label{fig:oneloop}
  \end{center}
\end{figure}

\begin{figure}[!htb]
  \begin{center}
    \includegraphics*{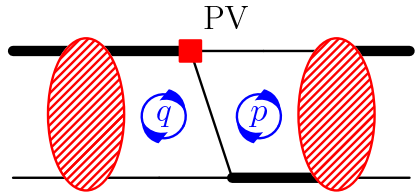}
    \caption{LO ``two-loop'' PV diagrams. ``Time-reversed'' (as defined in Fig.~\ref{fig:oneloop}) contributions not displayed.}
    \label{fig:twoloop}
  \end{center}
\end{figure}

The parity-violating $nd$ scattering amplitude receives contributions from
tree-level (Fig.~\ref{fig:tree}), ``one-loop'' (Fig.~\ref{fig:oneloop}), and
``two-loop'' diagrams (Fig.~\ref{fig:twoloop}). This nomenclature refers to the
number of loops that contain a parity-violating interaction; the strong
amplitudes receive contributions from an infinite series of
multi-loop diagrams. All graphs use the same interaction kernel, namely the
tree-level PV diagrams in the off-shell kinematics specified in
Fig.~\ref{fig:tree}.
These are in turn computed using the Lagrangians of Eqs.~(\ref{eq:PCLag}) and
(\ref{eq:PVLag}), and with the propagator $\calK$ of the exchanged nucleon in
Eq.~(\ref{eq:NucProp}) as well as the cluster-decomposition version of
the dibaryon propagator in Eq.~\eqref{eq:calD}. 
Amplitudes between two states with angular momenta $L$ and $L^\prime$ scale as
$k^{L+L^\prime}$ for small nucleon momenta, see e.g.~\cite[p.~381
Eq.~(157)]{GoldbergerWatson}. Since spin-rotation measurements are performed
with neutrons with momenta $k\ll 1\;\MeV$, only transitions between
$\mathrm{S}$ and $\mathrm{P}$ waves contribute at this order, i.e.~$L,L^\prime\in\{0;1\}$.
The projectors
necessary to filter out specific three-nucleon partial waves are constructed
and listed in App.~\ref{Sec:Projectors}.

While the PV Lagrangian contains five independent couplings, forward $Nd$ scattering  at low energies
only depends on three independent (isospin-dependent) linear combinations:
\begin{align}
  \calS_1 &= 3\daR +2\tau_3 \deR\;\;,\nonumber\\
\label{eq:calT}
  \calS_2 &= 3\daR -\tau_3 \deR\;\;,\\
  \calT &= 3\dbR +2\tau_3 \dcR\;\;.\nonumber
\end{align}
Since the $Nd$ system is an iso-doublet, the PV coupling $\ddR$ cannot
contribute. Here we are interested in scattering on a neutron, so only the $(22)$ component of the isospin matrix is needed, effectively replacing $\tau_3$ with $-1$.

The projected scattering amplitudes of Fig.~\ref{fig:tree} (a)
are: 
\begin{equation}\label{eq:TreeAmp}
  \begin{split}
    \ii\calM^{(a)}\left[\twoS \to \twoPone;q,p\right] &= \ii\frac{\sqrt{2}y
      M}{3pq}\left[2p Q_0(-\calE)+qQ_1(-\calE)\right]
  \begin{pmatrix}\calS_1&-\calT\\\calS_1&-\calT\end{pmatrix}
    , \\[1ex]
    i\calM^{(a)}\left[\twoPone \to \twoS;q,p\right] &= \ii\frac{\sqrt{2}y
      M}{3pq}\left[2p Q_1(-\calE)+qQ_0(-\calE)\right]
  \begin{pmatrix}\calS_1&-\calT\\\calS_1&-\calT\end{pmatrix}
    ,\\[1ex]
    \ii\calM^{(a)}\left[\twoS \to \fourPone;q,p\right] &= \ii\frac{4yM}{3pq}
    \left[2p Q_0(-\calE)+qQ_1(-\calE)\right] 
  \begin{pmatrix}\dis\frac{\calS_1-\calS_2}{3}&\calT\\0&0\end{pmatrix}
    ,\\[1ex]
    \ii\calM^{(a)}\left[\fourPone \to \twoS;q,p\right] &= \ii\frac{2y
      M}{3pq}\left[2p Q_1(-\calE)+qQ_0(-\calE)\right]
  \begin{pmatrix}\calS_2&0\\\calS_2&0\end{pmatrix}
    ,\\[1ex]
    \ii\calM^{(a)}\left[\fourS \to \twoPthree;q,p\right] &= \ii\frac{\sqrt{2}y
      M}{3pq}\left[2p Q_0(-\calE)+qQ_1(-\calE)\right]
  \begin{pmatrix}\calS_2&0\\\calS_2&0\end{pmatrix} Q^r{}_s, \\[1ex]
    \ii\calM^{(a)}\left[\twoPthree \to \fourS;q,p\right] &= \ii\frac{2\sqrt{2}y
      M}{3pq}\left[2p Q_1(-\calE)+qQ_0(-\calE)\right] 
  \begin{pmatrix}\dis\frac{\calS_1-\calS_2}{3}&\calT\\0&0\end{pmatrix}
    Q^r{}_s,\\[1ex]
    \ii\calM^{(a)}\left[\fourS \to \fourPthree;q,p\right] &= -\ii\frac{2\sqrt{10}y
      M}{3pq}\left[2p Q_0(-\calE)+qQ_1(-\calE)\right]
  \begin{pmatrix}\dis\frac{\calS_1-\calS_2}{3}&0\\0&0\end{pmatrix}
    Q^r{}_s, \\[1ex]
    \ii\calM^{(a)}\left[\fourPthree \to \fourS;q,p\right] &= -\ii\frac{2\sqrt{10}y
      M}{3pq}\left[2p Q_1(-\calE)+qQ_0(-\calE)\right] 
  \begin{pmatrix}\dis\frac{\calS_1-\calS_2}{3}&0\\0&0\end{pmatrix}
    Q^r{}_s,
  \end{split}
\end{equation}
with $E=\frac{3{\kv}^2}{4M}-\frac{\gamma_t^2}{M}+\ii\epsilon$ as before and
\begin{equation}
  \calE=\frac{p^2+q^2-ME}{pq}\;\;.
\end{equation}
The structure of these matrices in the cluster space of
Eq.~\eqref{eq:clusterdecomp} already appeared in
Ref.~\cite{Griesshammer:2010nd}.  The projector
$(Q^r{}_s)^\alpha{}_\beta=\delta^r_s\delta^\alpha_\beta-
\frac{1}{3}(\sigma^r\sigma_s)^\alpha{}_\beta$ onto the spin quartet is defined
in Eq.~\eqref{eq:Qij}. The index $\alpha$ ($\beta$) denotes the spin of the outgoing
(incoming) nucleon, and $r$ ($s$) is the spin component of the outgoing
(incoming) \threeS dibaryon. The diagrams of Fig.~\ref{fig:tree}(b) with the PV interactions on the lower line can be obtained from those of Fig.~\ref{fig:tree}(a) by
Hermitian conjugation in cluster, spin, and isospin space:
\begin{equation}
\label{eq:hermconjugate}
  \calM^{(b)}[X\to Y;q,p]^{r\alpha}{}_{s\beta}=\left(\calM^{(a)}[Y \to X;p,q]^{s\beta}{}_{r\alpha}\right)^\dagger\;\;,
\end{equation}
where $X$ ($Y$) denotes the partial wave of the incoming (outgoing) state and the spin indices are made explicit.  To
find the contributions of Fig.~\ref{fig:tree} to PV scattering,
choose the on-shell point $p=q=k$:
\begin{equation}
\label{eq:tree}
  \ii\calM_{\text{tree}}[X \to Y;k]=\ii\calM^{(a)}[X \to Y;k,k]+
  \ii\calM^{(b)}[X \to Y;k,k]
\end{equation} 
The ``one-loop'' and ``two-loop'' contributions of Figs.~\ref{fig:oneloop} and
\ref{fig:twoloop} are generated by convoluting the PV tree-level results with
the PC amplitudes of Sec.~\ref{Sec:PC}, following the calculation of
higher-order corrections in the PC sector in Ref.~\cite{Bedaque:1999vb}.

When the PC amplitude is attached to the left of the PV kernel in a ``one-loop''
diagram, as in Fig.~\ref{fig:oneloop}, the $q_0$ integration picks the nucleon
pole, $q_0=-\frac{q^2}{2M}+\ii\epsilon$, and the angular integration is
trivial, 
\begin{equation}
\label{eq:1loopleft}
  \ii\calM_{1\text{-loop, PC left}}^\text{LO}[X \to Y;k]=
  \sum\limits_{j=a,b}\int\limits_0^\Lambda \frac{\deint{}{q}q^2}{2\pi^2}\;
  \ii \calM^{(j)}\left[X \to Y;q,k\right]
  \ii\calD^\text{LO}(E;q)\;
  \ii t^\text{LO}[X;k,q]\;.
\end{equation}
Note that we choose the same UV regulator $\Lambda$ as in the integral
equations for the PC amplitudes,
Eqs.~(\ref{eq:quartetfaddeev}/\ref{eq:doubletfaddeev}).
When the PC amplitude is attached to the right, the amplitudes are obtained by
reading Fig.~\ref{fig:oneloop} as if the time direction were reversed:
\begin{equation}
\label{eq:1loopright}
  \ii\calM_{1\text{-loop, PC right}}^\text{LO}[X \to Y;k]=
  \sum\limits_{j=a,b}\int\limits_0^\Lambda \frac{\deint{}{q}q^2}{2\pi^2}\;
  \ii t^\text{LO}[Y;k,q]\;
  \ii\calD^\text{LO}(E;q)\;
  \ii \calM^{(j)}\left[X \to Y;k,q\right]\;,
\end{equation}
where we used that the
  PC amplitudes are time-reversal invariant, i.e.~identical when exchanging
  incoming and outgoing nucleon momenta, $t[X;p,q]=t[X;q,p]$.
The ``two-loop'' convolutions of Fig.~\ref{fig:twoloop} are
\begin{align}
\label{eq:2loop}
  &\ii\calM_{2\text{-loop}}^\text{LO}[X \to Y;k]=\\
  &  \sum\limits_{j=a,b}\int\limits_0^\Lambda \frac{\deint{}{q}q^2}{2\pi^2}
  \int\limits_0^\Lambda \frac{\deint{}{p}p^2}{2\pi^2}\; \ii t^\text{LO}[Y;k,p]\;
  \ii\calD^\text{LO}(E;p)\; \ii \calM^{(j)}\left[X \to Y;q,p\right]
  \ii\calD^\text{LO}(E;q)\; \ii t^\text{LO}[X;k,q]\;\;.\notag
\end{align}
We numerically solve the integral equations for the PC amplitudes $t$,
Eqs.~(\ref{eq:quartetfaddeev}/\ref{eq:doubletfaddeev}), by the
Hetherington-Schick method~\cite{HetheringtonSchick} in a Mathematica code,
with $\Lambda$ a hard cutoff corresponding to a step function, as detailed in
Ref.~\cite{Griesshammer:2004pe}. The convolutions for the PV amplitudes
Eqs.~(\ref{eq:1loopleft}/\ref{eq:1loopright}/\ref{eq:2loop}) are performed
numerically, using again a Mathematica code.  The numerical error of $\lesssim
0.1\%$ of these procedures is negligible relative to the systematic
uncertainties discussed in Sec.~\ref{sec:ndresult}.

Nucleon-deuteron scattering corresponds to the (11) element of the
cluster matrix.
Multiplication with the LO wave-function normalization factor
$\sqrt{\calZ_t^\text{LO}}$, Eq.~\eqref{eq:calZLO}, for each external deuteron
leg results in the renormalized, physical scattering amplitude
$\calM^\text{LO}_\text{R}$ between partial waves $X$ and $Y$ at nucleon
momentum $k$:
\begin{equation}
  \label{eq:renampLO}
  \begin{split}
  \ii\calM^\text{LO}_\text{R}[X \to
  Y;k]=&\left(\sqrt{\calZ_t^\text{LO}},0\right)\Big( 
    \ii\calM_{\text{tree}}[X \to Y;k]+
    \ii\calM_{1\text{-loop, PC right}}^\text{LO}[X \to Y;k]\\&
    +\ii\calM_{1\text{-loop, PC left}}^\text{LO}[X \to Y;k]+
    \ii\calM_{2\text{-loop}}^\text{LO}
    [X \to Y;k]\Big)\left.\binom{\sqrt{\calZ_t^\text{LO}}}{0}\right|_{\tau_3\to -1}\;.
  \end{split}
\end{equation}

\subsection{$nd$ partial-wave amplitudes at next-to-leading order}
\label{sec:ndNLO}

No new PV operators contribute to the NLO amplitudes since higher-order PV interactions are suppressed by at least $Q^2$ as described in Sec.~\ref{Sec:PVLag}. 
There are three types of NLO corrections. The first type is given by
diagrams with one insertion of an effective-range correction
to the dibaryon propagators proportional to $c_{0s/t}$, see Eq.~\eqref{eq:c0}. At
this order, the momentum-independent PC 3NI
parameter $H_0(\Lambda)$ requires an additive adjustment,
called $H_0^\text{NLO}(\Lambda)$, to recapture the low-energy 3N observable
that initially fixed $H_0(\Lambda)$ at leading order. This results
in a second type of NLO correction to the $nd$ system: diagrams with
$H_0^\text{NLO}(\Lambda)$ inserted once between LO amplitudes.
The third type of corrections comes from the change in the dibaryon
wave-function renormalization and is taken into account by replacing
$\calZ_{t}^\text{LO}$ in the LO PV amplitude of Eq.~\eqref{eq:renampLO} by
$\calZ_{t}^\text{LO+NLO}$, see Eq.~\eqref{eq:calZNLO}.

\begin{figure}[!htbp]
  \begin{center}
    \includegraphics*{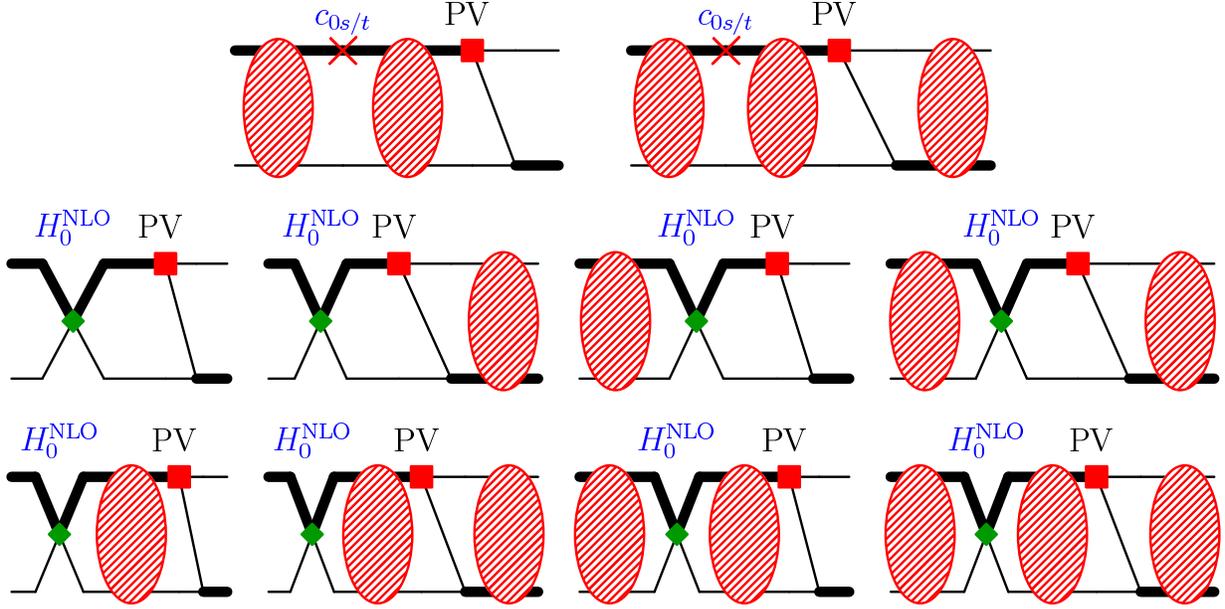}
    \caption{Class-I diagrams.  Diagrams with the PV
      vertex on the lower line as well as ``time-reversed'' contributions not
      displayed.}
    \label{fig:PV3NstrictNLOdifferentloop}
  \end{center}
\end{figure}

\begin{figure}[!htbp]
  \begin{center}
    \includegraphics*{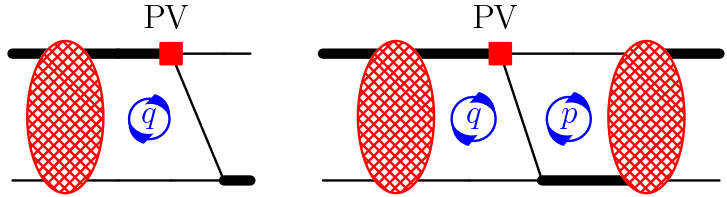}
    \caption{Reformulation of the class-I diagrams of
      Fig.~\ref{fig:PV3NstrictNLOdifferentloop} using the amplitudes $t^\text{LO+NLO}$ of Fig.~\ref{fig:3NscatteringNLO}. This
      also replaces the LO PV contributions of Figs.~\ref{fig:oneloop}
      and \ref{fig:twoloop}. Diagrams with the PV vertex on the lower line as well
      as ``time-reversed'' contributions not displayed.}
    \label{fig:PV3N-NLOdifferentloop}
  \end{center}
\end{figure}

In a strictly perturbative approach, there are two classes of diagrams at NLO. The class-I diagrams of Fig.~\ref{fig:PV3NstrictNLOdifferentloop} are generated when the PC LO amplitudes $t^\text{LO}[X;k,q]$ are replaced with the partially-resummed amplitudes $t^\text{LO+NLO}[X;k,q]$ in the topologies of the LO PV ``one-'' and ``two-loop'' diagrams of Figs.~\ref{fig:oneloop} and~\ref{fig:twoloop}, i.e.~$t^\text{LO}[X;k,q]\to t^\text{LO+NLO}[X;k,q]$ at each occurrence in Eqs.~(\ref{eq:1loopleft}/\ref{eq:1loopright}/\ref{eq:2loop}), see Fig.~\ref{fig:PV3N-NLOdifferentloop}. 
In addition to the NLO class-I diagrams, the amplitudes obtained from Fig.~\ref{fig:PV3N-NLOdifferentloop} also contain all LO contributions as well as a partial resummation of higher-order effects. In analogy to the notation at LO, the resulting contributions are called $\calM_{1\text{-loop, PC left}}^\text{LO+NLO, class-I}[X \to Y;k]$, $\calM_{1\text{-loop, PC right}}^\text{LO+NLO, class-I}[X \to Y;k]$, and $\calM_{2\text{-loop}}^\text{LO+NLO, class-I}[X\to Y;k]$. This approach has computational advantages since it avoids diagrams with up to four numerical integrations and convolutions with parity-conserving LO full-off-shell amplitudes $t^\text{LO}[{}^{2S+1}L_{J};p,q]$, $p,q\not=k$. In addition, the diagrams in class I contain divergences, studied by Refs.~\cite{Hammer:2001gh,Griesshammer:2010nd,Ji:2011qg}, that in a strictly perturbative calculation are cancelled by corresponding contributions in class-II diagrams, to be discussed shortly.  These divergences are challenging to treat numerically, but are avoided in the approach used here. The amplitudes corresponding to the diagrams of Fig.~\ref{fig:PV3N-NLOdifferentloop} are individually renormalized; each diagram approaches a unique and finite limit as the cutoff is removed, $\Lambda\to\infty$.

\begin{figure}[!htbp]
  \begin{center}
    \includegraphics*{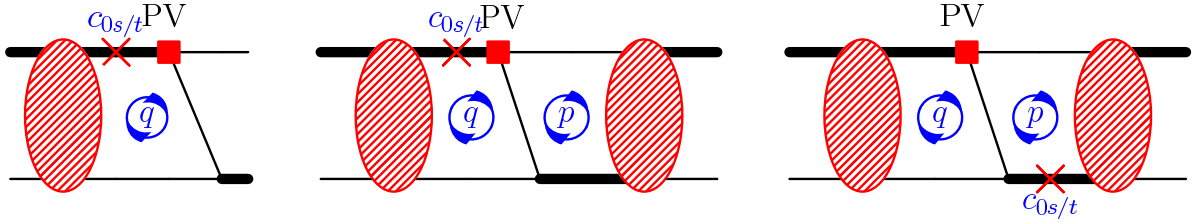}
    \caption{Class-II diagrams. As discussed in the text,
      the PC amplitudes are taken as $t^\text{LO+NLO}$ for \twoS and \fourS,
      and as $t^\text{LO}$ for \wave{2}{P}{J} and \wave{4}{P}{J}.  Diagrams
        with the PV vertex on the lower line as well as ``time-reversed''
      contributions not displayed.}
    \label{fig:PV3NstrictNLOsameloop}
  \end{center}
\end{figure}

Diagrams that are not generated by the replacement $t^\text{LO}[X;k,q]\to t^\text{LO+NLO}[X;k,q]$ in Figs.~\ref{fig:oneloop} and~\ref{fig:twoloop} are referred to as ``class-II'' diagrams and are shown in Fig.~\ref{fig:PV3NstrictNLOsameloop}.
The corresponding amplitudes are given by
\begin{align}
\label{eq:1loopinsertionleft}
  &\ii\calM_{1\text{-loop, PC left}}^\text{NLO, class-II}[X \to Y;k]=
  \sum\limits_{j=a,b}\int\limits_0^\Lambda \frac{\deint{}{q}q^2}{2\pi^2}\;
  \ii \calM^{(j)}\left[X \to Y;q,k\right]\;  \ii\calD^\text{NLO}(E;q)\;
  \ii t[X;k,q]\;,\\
\label{eq:1loopinsertionright}
  &\ii\calM_{1\text{-loop, PC right}}^\text{NLO, class-II}[X \to Y;k]=
  \sum\limits_{j=a,b}\int\limits_0^\Lambda \frac{\deint{}{q}q^2}{2\pi^2}\;
  \ii t[Y;k,q]\; \ii\calD^\text{NLO}(E;q)\;
  \ii \calM^{(j)}\left[X \to Y;k,q\right]\;,\\
\label{eq:2loopinsertion}
&\ii\calM_{2\text{-loop}}^\text{NLO, class-II}[X \to Y;k]=\\
&\quad\quad\quad\quad \sum\limits_{j=a,b}\int\limits_0^\Lambda
\frac{\deint{}{q}q^2}{2\pi^2} \int\limits_0^\Lambda
\frac{\deint{}{p}p^2}{2\pi^2}\; \ii t[Y;k,p]\Big[\ii\calD^\text{LO}(E;p)\;
  \ii \calM^{(j)}\left[X \to Y;q,p\right]
  \ii\calD^\text{NLO}(E;q)\nonumber\\
  &\quad\quad\quad\quad\quad\quad\quad\quad\quad\quad+\ii\calD^\text{NLO}(E;p)\; \ii
  \calM^{(j)}\left[X \to Y;q,p\right] \ii\calD^\text{LO}(E;q)\Big] \ii
t[X;k,q]\;\;.\notag
\end{align}
Since the kernel of these convolutions is now already NLO, it appears that the LO PC amplitudes $t^\text{LO}[X;k,q]$ can be used directly. This is the
strategy implemented for the PC $\text{P}$-wave amplitudes. 

For a convolution of the class-II diagrams involving the LO PC \twoS-wave amplitude the situation is more complicated.  It was demonstrated on general grounds in Ref.~\cite{Griesshammer:2010nd}, and confirmed numerically here that when using $t^\text{LO}$ the ``one-loop'' amplitudes diverge logarithmically, and the ``two-loop'' amplitudes diverge as $q^{0.23\dots}$; see App.~\ref{app:insertions}. These are \emph{not} divergences that are removed by additional parity-violating 3NIs at NLO \cite{Griesshammer:2010nd}. In a strictly perturbative calculation in the PC sector, the 3NI $H_0^\text{NLO}$ absorbs a linear divergence generated by insertions of the
effective-range term \cite{Hammer:2001gh,Ji:2011qg}. It is this linear divergence in $H_0^\text{NLO}$ which, when inserted next to a PV interaction (see class-I diagrams in Fig.~\ref{fig:PV3NstrictNLOdifferentloop}), would renormalize the divergence in the class-II contributions. However, in the partially-resummed formalism used here for the class-I diagrams $H_0^\text{LO+NLO}$ does not diverge linearly as $\Lambda\to\infty$, see Fig.~\ref{fig:H0dependence}, since  the high-off-shell momentum part of the scattering equation, and hence of the amplitude, becomes softer.
Therefore, the treatment of class-I diagrams using the partial resummation technique, see Fig.~\ref{fig:PV3N-NLOdifferentloop},
removes those divergences from the class-I diagrams. This in turn means that the class-I diagrams do not renormalize the divergent class-II contributions, so that the divergences in the class-II diagrams require separate renormalization. By using the renormalized (partially-resummed) PC amplitude $t^\text{LO+NLO}[\twoS;k,q]$ for the \twoS channel the UV behavior of the class-II diagrams is changed and no divergences appear. As demonstrated below in Fig.~\ref{fig:partialwaveresults-nd-2S}, this leads to renormalized, cutoff-independent PV amplitudes. We therefore choose the renormalized PC amplitude $t^\text{LO+NLO}[\twoS;k,q]$ for this channel.

We also choose the LO+NLO version for the PC \fourS-wave amplitude in the
class-II diagrams. This is not required by renormalization, as the
loop-integrations in
Eqs.~(\ref{eq:1loopinsertionleft}/\ref{eq:1loopinsertionright}/\ref{eq:2loopinsertion})
converge well in this channel, see again Ref.~\cite{Griesshammer:2010nd}.
However, neutron spin-rotation in deuterium is particularly sensitive to
the scattering lengths of the $nd$ system since it is a
process essentially at zero kinetic energy. It is therefore important to reproduce the experimental values of the scattering lengths. In particular, the \fourS scattering
length is a factor of $10$ larger than the \twoS one, so it will most likely
dominate spin-rotation observables. 
At LO in the Z-parameterization used
here, the \fourS scattering length of $5.1\;\fm$ would require
a 25 percent correction to achieve the experimental value
of $[6.35\pm0.02]\;\fm$~\cite{doublet_sca}, while
the NLO amplitude leads to $6.4\;\fm$, in close agreement with the experimental value, see~Ref.~\cite{Griesshammer:2004pe}. We
therefore use the NLO PC amplitude $t^\text{LO+NLO}[\fourS;k,q]$ for this channel. This choice does not violate the power
counting of the theory and yet improves the utility of our result.

To summarize, the PC amplitudes used in
Eqs.~(\ref{eq:1loopinsertionleft}/\ref{eq:1loopinsertionright}/\ref{eq:2loopinsertion}) are:
\begin{equation}
 \label{eq:insertionchoice}
  t[X;k,q]=\left\{\begin{array}{ll}
      t^\text{LO+NLO}[X;k,q]& \mbox{ for } X\in\{\twoS;\fourS\}\;,\\
      t^\text{LO}[X;k,q]& \mbox{ for } X\in\{\wave{2}{P}{};\wave{4}{P}{}\}\;.
    \end{array}\right.
\end{equation}
Further details and comparisons to other choices are discussed in
App.~\ref{app:insertions}.

The renormalized LO+NLO PV amplitudes are obtained by summing and
multiplying with the LO+NLO wave-function renormalization:
\begin{align}
  \label{eq:renampNLO}
    \ii\calM^\text{LO+NLO}_\text{R}&[X \to
    Y;k]=\left(\sqrt{\calZ_t^\text{LO+NLO}},0\right)\Big(
    \ii\calM_{\text{tree}}[X \to Y;k]\\&+
    \ii\calM_{1\text{-loop, PC left}}^\text{LO+NLO, class-I}[X \to Y;k]+ 
    \ii\calM_{1\text{-loop, PC left}}^\text{NLO, class-II}[X \to Y;k]\nonumber\\&+
    \ii\calM_{1\text{-loop, PC right}}^\text{LO+NLO, class-I}[X \to Y;k]+ 
    \ii\calM_{1\text{-loop, PC right}}^\text{NLO, class-II}[X \to Y;k]\nonumber\\&+
    \ii\calM_{2\text{-loop}}^\text{LO+NLO, class-I}[X \to Y;k]+
    \ii\calM_{2\text{-loop}}^\text{NLO, class-II}[X \to Y;k]
\Big)\binom{\sqrt{\calZ_t^\text{LO+NLO}}}{0}\nonumber
\end{align}
Finally, we reiterate that these amplitudes are complete up to NLO, but also
contain some higher-order contributions. The PC amplitudes $t^\text{LO+NLO}$
used are partially resummed. In addition, we choose to multiply the entire
LO+NLO amplitude by the LO+NLO wave-function renormalization, rather than the
LO PV amplitude $\calM^\text{LO}_\text{R}$ by $\calZ^\text{LO+NLO}$ and the
NLO correction $(\calM^\text{LO+NLO}_\text{R}-\calM^\text{LO}_\text{R})$ only
by $\calZ^\text{LO}$. It has been demonstrated repeatedly that this speeds up
convergence since the larger-than-usual wave-function renormalization
correction at NLO is expected to be the dominant correction to insertions of
higher-dimension operators; see
e.g.~Refs.~\cite{Phillips:1999hh,Beane:2000fx,Griesshammer:2004pe}.

The computational effort can be halved by taking advantage of the fact that
the individual tree-level amplitudes $\calM^{(a)}$ and $\calM^{(b)}$ are
Hermitian conjugates, Eq.~\eqref{eq:hermconjugate}, and that the PC amplitudes
are time-reversal invariant, $t[X;p,q]=t[X;q,p]$. The amplitudes of partial
waves $X \to Y$ and $Y \to X$ are therefore related:
\begin{align}
 \calM_{1\text{-loop, PC right}}[X \to Y;k]&=\left(\calM_{1\text{-loop,
     PC left}}[Y \to X;k]\right)^\dagger\;,\\
 \calM_{1\text{-loop}}[X \to Y;k]&=\left(\calM_{1\text{-loop}}[Y \to
   X;k]\right)^\dagger\;,\\ 
 \calM_{2\text{-loop}}[X \to Y;k]&=\left(\calM_{2\text{-loop}}[Y \to
   X;k]\right)^\dagger\;,\\
\label{eq:conjugateamps}
\calM_\text{R}[X \to Y;k]&=\calM_\text{R}[Y \to X;k]\;.
\end{align}

\subsection{Cutoff independence of partial-wave amplitudes}
\label{Sec:CutoffInd}

Numerically, it is difficult to perform some of the integrations at very low
energies. However, between $k=0.01\;\MeV$ and
$10\;\MeV$ the amplitudes of the $nd$ system deviate from the linear relation
only at the percent-level. 
For the spin rotation angle, the relevant quantity is $\calM_\text{R}/k$, see Eq.~\eqref{eq:spinrotation}. The difference in $\calM_\text{R}/k$ at $k=1\;\MeV$ and
$0.01\;\MeV$ is less than $0.3\%$. All results quoted are for $k=1\;\MeV$.

\begin{figure}[!htbp]
  \begin{center}
    \includegraphics*[width=\linewidth]{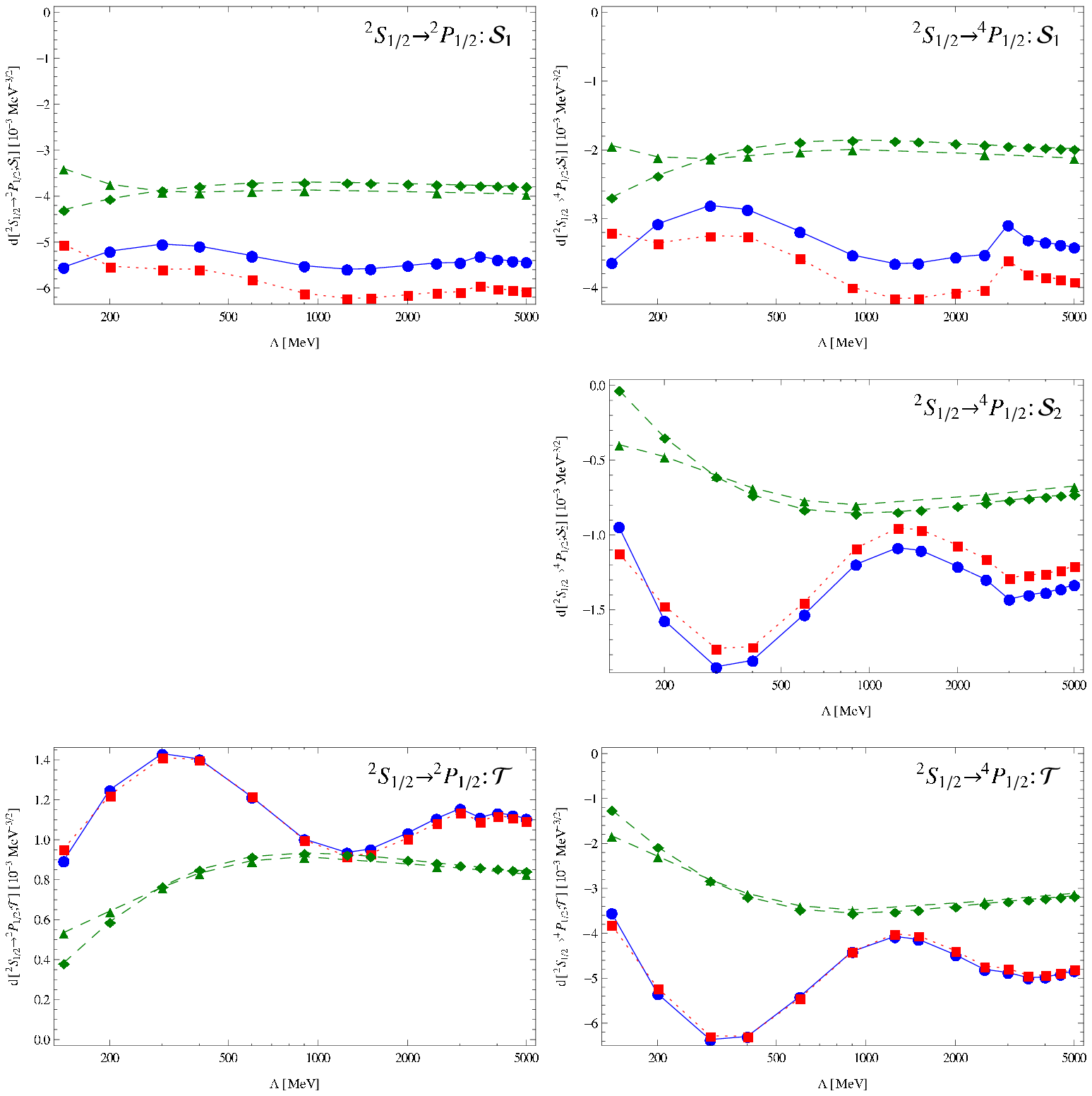}
    \caption{Cutoff dependence of the functions $d[\twoS\to
      Y;\text{coupling}](\Lambda)$, Eq.~\eqref{eq:ndPWdecomposed}. Columns (left to right): partial waves
      $\twoS\to\wave{2}{P}{\half}$, $\twoS\to\wave{4}{P}{\half}$; rows (top to
      bottom): coefficients of $\calS_1$, $\calS_2$, $\calT$.  \protect\gdiamond 
      (dashed lines): LO, 3NI $H_0$ determined from $nd$ \twoS-scatt.~length;
      \gtriangle (dashed lines): LO, 3NI from triton binding energy; \bdisc
      (solid lines): NLO, 3NI from \twoS-scatt.~length; \protect\rsquare
      (dotted lines): NLO, 3NI from triton binding energy. The linear
      extrapolations are only meant to guide the eye.  Notice the
      different scales on the vertical axes.  Amplitudes which are identically
      zero are not displayed. }
    \label{fig:partialwaveresults-nd-2S}
  \end{center}
\end{figure}

\begin{figure}[!htbp]
  \begin{center}
    \includegraphics*[width=0.96\linewidth]{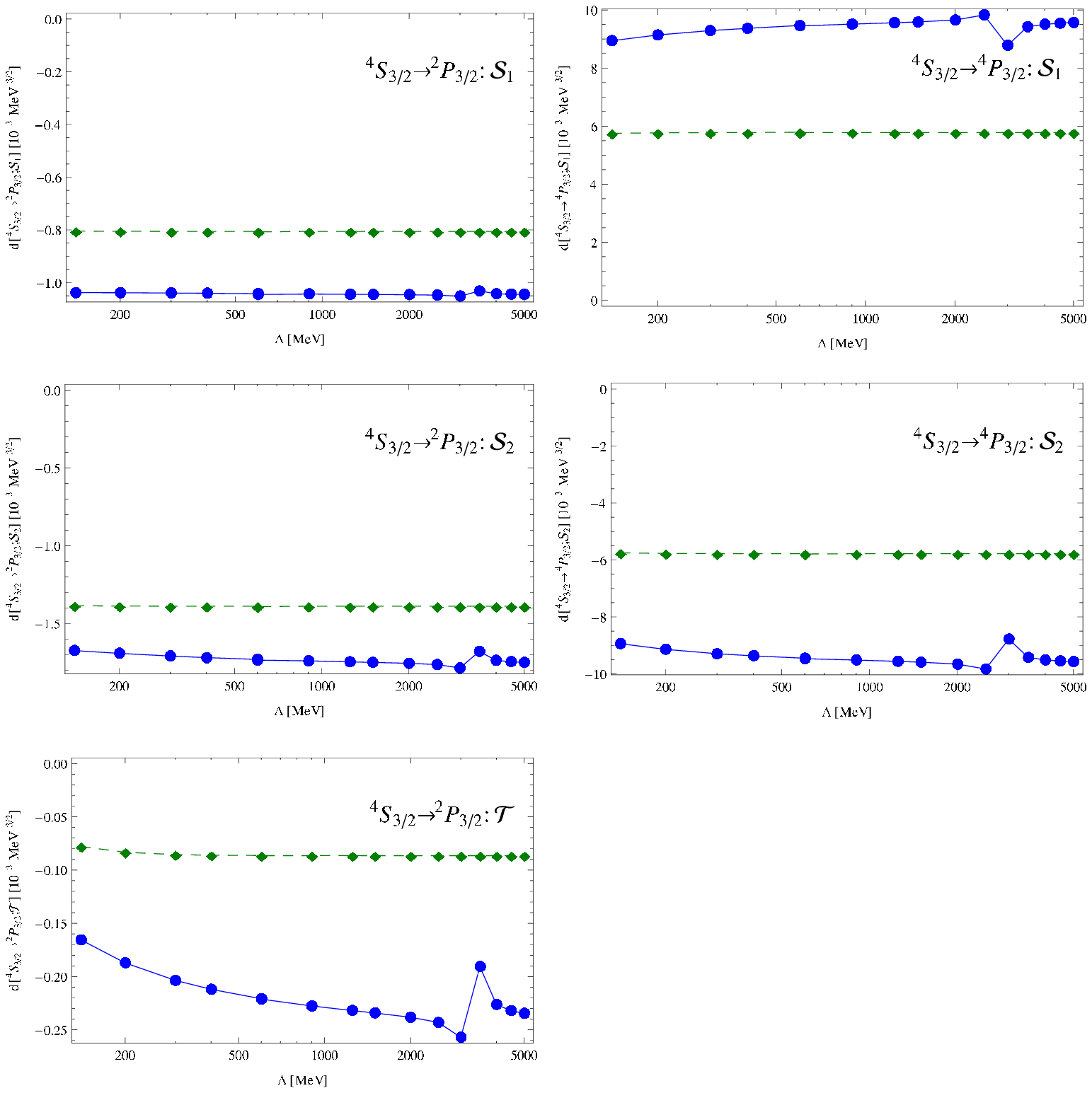}
    \caption{Cutoff dependence of the functions $d[\fourS\to
      Y;\text{coupling}](\Lambda)$,
      Eq.~\eqref{eq:ndPWdecomposed}. Columns (left to right): partial waves
      $\fourS\to\wave{2}{P}{\frac{3}{2}}$, $\fourS\to\wave{4}{P}{\frac{3}{2}}$; rows (top to
      bottom): coefficients of $\calS_1$, $\calS_2$, $\calT$. \protect\gdiamond 
      (dashed lines): LO; \bdisc
      (solid lines): NLO. Comments as in Fig.~\ref{fig:partialwaveresults-nd-2S}.}
    \label{fig:partialwaveresults-nd-4S}
  \end{center}
\end{figure}

For the analysis of the cutoff dependence of the partial-wave amplitudes, we decompose them
in terms of the PV interactions $\calS_1$, $\calS_2$, and $\calT$ of Eq.~\ref{eq:calT}:
\begin{align}
  \label{eq:ndPWdecomposed}
  \frac{\Re[\calM_\text{R}[X \to Y;k]]}{k} = &  d[X \to Y;\calS_1](\Lambda)\;\calS_1\\
  &+ d[X \to Y;\calS_2](\Lambda)\;\calS_2+d[X \to
  Y;\calT](\Lambda)\;\calT\;\;.\nonumber
\end{align}
In Figs.~\ref{fig:partialwaveresults-nd-2S}
and~\ref{fig:partialwaveresults-nd-4S}, the dependence of the functions
$d$ on the cutoff $\Lambda$ used in the PC 3N integral equations
Eqs.~(\ref{eq:quartetfaddeev}/\ref{eq:doubletfaddeev}) and in the convolutions
of Secs.~\ref{sec:ndLO} and~\ref{sec:ndNLO} is shown for each partial-wave and
order.
Since a unique, finite limit exists as $\Lambda\to\infty$, the result is
properly renormalized in each partial wave. These results confirm the theoretical findings of Ref.~\cite{Griesshammer:2010nd} that no PV 3NI is required at LO and NLO. Comparison with Fig.~\ref{fig:H0dependence} also
shows that at both LO and NLO the physical amplitudes are smooth even where the
3NI $H_0$ diverges.

A small kink in the cutoff dependence of the partial wave amplitudes is seen at NLO for $\Lambda\approx 3000\;\MeV$. It is evident in all partial waves, and so is not related to the renormalization of the 3NI $H_0$ of the \twoS-wave. The phenomenon appears only at cutoffs well beyond the breakdown scale of 200 MeV and hence does not have any impact on our final result.

\subsection{Translating partial-wave amplitudes into neutron-spin rotation
  predictions}
\label{Sec:spinrot-formulae}

For computational convenience, the calculations so far have been performed in
a partial-wave basis. However, in order to obtain the spin rotation angle, we
need to determine the forward scattering amplitude for specific neutron
helicity states. Recall that in the kinematics of
Sec.~\ref{Sec:PC}, the incident and outgoing nucleons have momentum $-\kv = - k \vec{e}_z$, so
that an incoming neutron $N^{a\alpha}(-\kv)$ with positive helicity corresponds to choosing the
spin and isospin components $\alpha=2,\;a=2$. The relation to the results in the partial-wave basis is obtained by inserting a complete set of projection operators:
\begin{equation}
  \langle d_j N_{b\beta}(-\kv)\lvert \calM \rvert d^i N^{a\alpha}(-\kv)\rangle 
  = \sum_{XY}\langle d_j N_{b\beta}(-\kv)| Y \rangle \langle Y \lvert \calM
  \rvert X \rangle \langle X | d^i N^{a\alpha}(-\kv)\rangle\;\;, 
\end{equation}
where $\langle Y \lvert \calM \rvert X\rangle=\calM_\text{R}[X\to Y]$ are the
partial-wave amplitudes calculated above. The sum over $X,Y$ stands for
all partial waves including spin and isospin polarizations.

Using the three-body projectors $\calP[X]$ constructed in
App.~\ref{Sec:Projectors} from Eq.~\eqref{eq:projectonsource}, the
partial-wave projected matrix elements of the neutron-deuteron state are
\begin{equation}
  \langle X=\left({}^{2S+1}{L}_{J}; b\beta\{M\}\right) | d^i
  N^{a\alpha}(-\kv)\rangle = 
  \left[\left(\calP^{\{M\}}{}_{i}[{}^{2S+1}{L}_{J}]
    \right)^{b\beta}{}_{a\alpha}\right]_{11}\;\;,
\end{equation}
with appropriate vector and spinor indices $\{M\}$. The outer bracket
$[.]_{11}$ indicates that only the $(11)$-entry of the cluster-decomposition
matrix is physically allowed. As an example, the projection onto the
\wave{4}{P}{\half}-wave is
\begin{equation}
  \langle \wave{4}{P}{\half}  | d_i N^{a\alpha}(-k\ev_k)\rangle =
  \sqrt{\frac{3}{2}} \left(Q^3{}_i\right)^\beta{}_\alpha\; \delta^b_a\;\;, 
\end{equation}
where the indices $\beta,\;b$ are needed to specify the spinor and isospinor
magnetic quantum numbers of the \wave{4}{P}{\half} state, and $\ev_k=(0,0,1)$
for forward-scattering.  The
positive-helicity neutron-deuteron amplitude is then: 
\begin{equation}
\label{eq:poshelamp}
\calM_+=\sum_{XY} \calM_\text{R}[X\to Y]\;\frac{1}{3}\sum_{i\in\{0;\pm1\}}
\left[\left(\left(\calP_{\{M\}}{}^{iA}[Y]\right)^\dagger\calP^{\{M\}}{}_{iA}[X]
\right)_{\beta=2,\alpha=2;a=2,b=2}\right]_{11}\;\;.
\end{equation}
Note that the iso-vector index $A$ is irrelevant for the
final $(11)$-component of the cluster-decomposition matrix. This formula
can also be understood as a decomposition into the reduced matrix elements,
represented by the amplitudes $\calM_\text{R}[X\to Y]$ calculated above, and
the equivalent of Clebsch-Gordan coefficients weighting the various partial
waves.  The negative-helicity amplitude is $\calM_-=-\calM_+$. Using the partial-wave projectors of App.~\ref{Sec:Projectors}
leads to our final result for the neutron spin-rotation in deuterium from
Eq.~\eqref{eq:spinrotation} with $\mu=2M/3$ for the reduced mass: 
\begin{align}
\label{eq:poshelampparameterised}
\frac{1}{\rho}\;\frac{\dd \phi_\text{PV}^{nd}}{\dd l} =\frac{2M}{6k}\;\frac{4}{9}\;
\Re\bigg[&\calM_\text{R}[\twoS\to\wave{2}{P}{\half};k]
-2\sqrt{2}\calM_\text{R}[\twoS\to\wave{4}{P}{\half};k]\\&
-4\calM_\text{R}[\fourS\to\wave{2}{P}{\frac{3}{2}};k]
-2\sqrt{5}\calM_\text{R}[\fourS\to\wave{4}{P}{\frac{3}{2}};k]\bigg]\;,\nonumber
\end{align}
where the amplitudes for P to S wave transitions are taken into account by using the relations of Eq.~\eqref{eq:conjugateamps}.

\subsection{Numerical $nd$ spin rotation result and error estimates}
\label{sec:ndresult}

\begin{figure}[!htbp]
  \begin{center}
    \includegraphics*[width=0.5\linewidth]{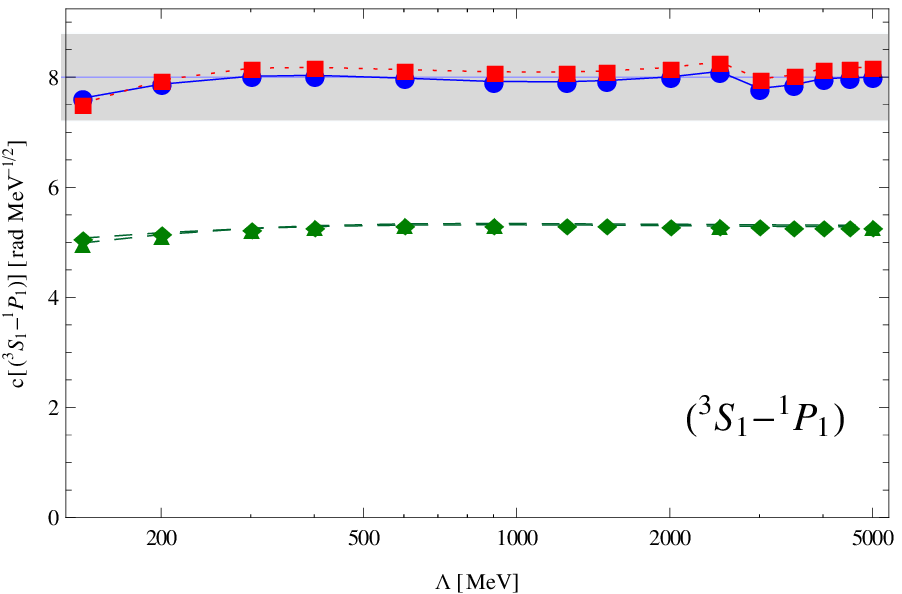}\\
    \includegraphics*[width=0.5\linewidth]{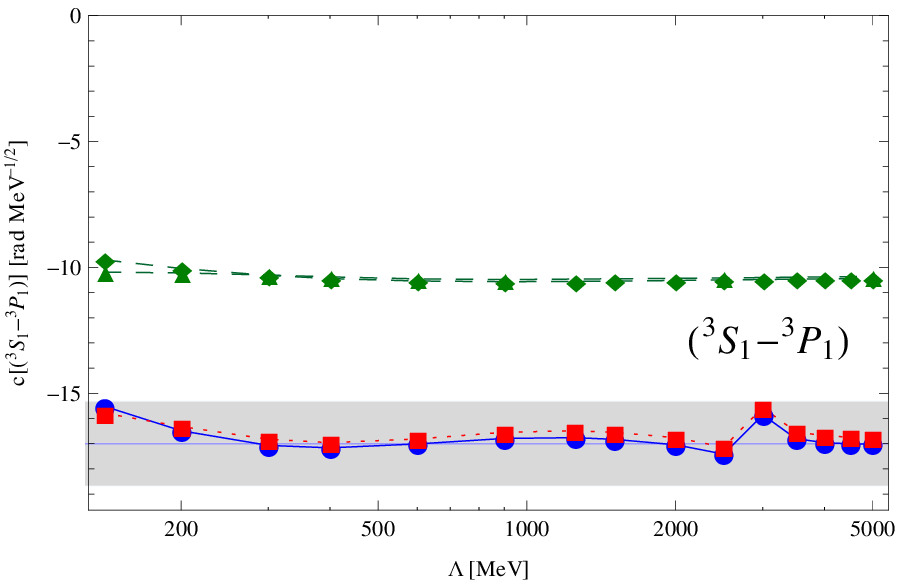}\\
    \includegraphics*[width=0.5\linewidth]{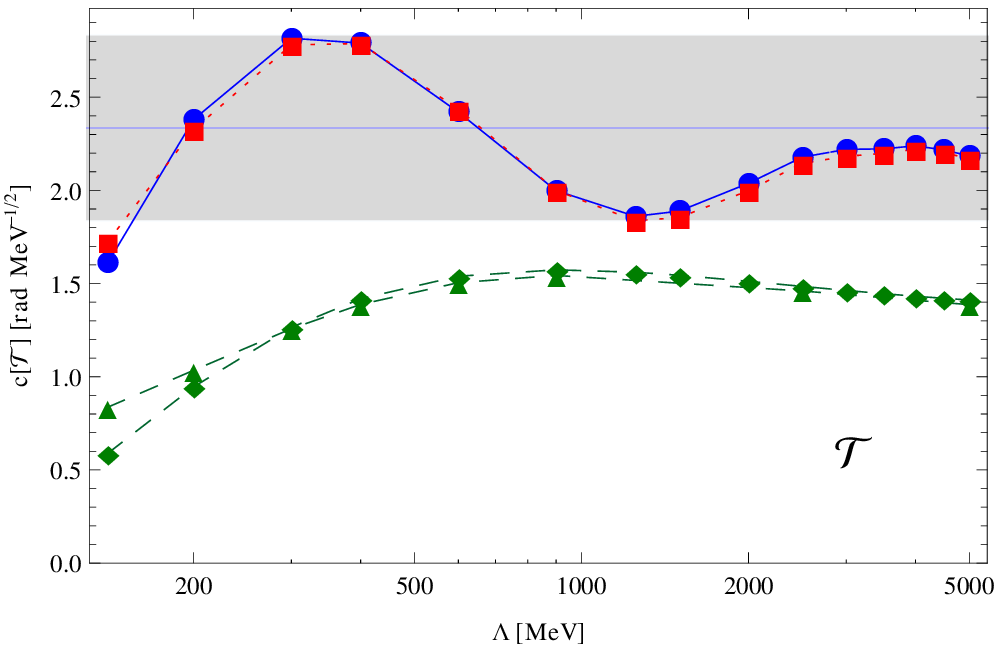}
    \caption{Cutoff dependence of the functions  $c[(X-Y)](\Lambda)$ for
      the neutron spin-rotation angle in deuterium,
      \eqref{eq:ndspinrotdecomposed}. Notation as in
      Fig.~\ref{fig:partialwaveresults-nd-2S}. Grey band: estimated
      theoretical uncertainties as described in Eq.~\eqref{eq:ndspinrotresult}.
      The linear extrapolations are only meant
      to guide the eye. Notice the different scales on the vertical axes.}
    \label{fig:spinrotresults-nd}
  \end{center}
\end{figure}

For the detailed discussion of theoretical errors below, we decompose the spin-rotation result of
Eq.~\eqref{eq:poshelampparameterised} in terms of  functions $c[(X-Y)](\Lambda)$
which multiply the PV interactions:
\begin{equation}
  \label{eq:ndspinrotdecomposed}
  \frac{1}{\rho}\;\frac{\dd \phi_\text{PV}^{nd}}{\dd l}
 =c[(^3 \! S_1-^1 \! P_1)](\Lambda)\;\daR+\tau_3\;c[(^3 \! S_1-^3 \! P_1)](\Lambda)\;\deR+c[\calT](\Lambda)\;\calT \;
\end{equation}  
The isospin matrix $\tau_3$ is replaced by $-1$ for neutron spin-rotation, and
$ \calT = 3\dbR +2\tau_3 \dcR$, see Eq.~\eqref{eq:calT}. 
Figure~\ref{fig:spinrotresults-nd} shows the cutoff dependence of the
functions  $c[(X-Y)](\Lambda)$.
As expected from the discussion of the cutoff dependence of the partial-wave amplitudes, the result for the spin-rotation angle is also properly renormalized.

The final result for neutron spin-rotation in deuterium is: 
\begin{align}
  \label{eq:ndspinrotresult}
  \frac{1}{\rho}\;\frac{\dd \phi_\text{PV}^{nd}}{\dd l}
  =&\;\Big([8.0\pm0.8]\;\daR\;+\;[17.0\pm1.7]\;\deR
  \\
  & +\;[2.3\pm0.5]\; \left(3\dbR - 2 \dcR\right) \Big)\text{rad}\;\MeV^{-\half}\;,\nonumber
\end{align}  
where the estimate of the residual theoretical uncertainties is justified below. It will turn out that this estimate is rather conservative;
drawing from the experience in calculations of PC 3N observables using the
$Z$-parameterization, theoretical systematic errors might be estimated to be on the
order of $3$\%, see e.g.~\cite{Griesshammer:2004pe,Kirscher:2009aj}. For now, we note that the theoretical errors of $10\%$ to $20\%$ are comparable to the statistical and systematic errors expected of the most ambitious
planned experiments.

We use three methods to estimate theoretical uncertainties, with the \EFTNoPion parameter $Q\approx\frac{1}{3}$ as a
conservative value for typical momenta in the $Nd$ system on the order of $\gamma_t$ (see Ref.~\cite{Griesshammer:2004pe}).
\begin{itemize}
\item[(i)] At NLO, higher orders should contribute corrections of order
  $Q^2\approx0.1$ to the total result.

\item[(ii)] Since \EFTNoPion  is not valid at high momentum, a residual dependence of observables on
  momentum modes above the breakdown scale $\bar{\Lambda}\sim200\;\MeV$ is
  an indicator of the size of higher-order corrections.  Varying
  the momentum cutoff $\Lambda$ in the Faddeev equations and convolutions of
  the 3N system from the breakdown scale to higher values provides an estimate of \NXLO{2}
  effects. The parameters $g^{(X-Y)}$ of the
  PV Lagrangian Eq.~\eqref{eq:PVLag} are independent of the regularization
  scale $\Lambda$. Therefore,  the $c[(X-Y)](\Lambda)$ of the
  spin-rotation results of Eq.~\eqref{eq:ndspinrotdecomposed} must also 
  be cutoff-independent up to higher-order effects. In particular, this
  implies that these coefficients should approach a unique, finite limit as
  the cutoff is removed. As in a variety of previous calculations in the 3N
  system, we consider cutoff-variations from $\Lambda=200\;\MeV$ to
  $5000\;\MeV$, see e.g.~Refs.~\cite{Griesshammer:2004pe,Bedaque:2002yg}.  
  
\item[(iii)] Finally, the size of higher-order effects can be assessed by
  using different low-energy data to determine the PC parameters of \EFTNoPion.
  Since we chose Z-parameterization precisely because of its well-established
  improved convergence, we do not vary parameters of the $NN$ system, e.g.~by
  replacing $(Z_{s/t}-1)$ at NLO by the effective ranges. On the other hand,
  the strength of the PC 3NI $H_0$ can for example be determined from the $nd$
  scattering length of the \twoS-wave, or from the triton binding energy. The
  difference between both approaches is again a measure of \NXLO{2} effects,
  i.e.~expected to be on the order of $Q^2\approx0.1$. 
  
\end{itemize}
Note that methods (ii) and (iii) do not apply to the 2N system.
We will base our error-estimate for the $nd$ spin-rotation
coefficients on the most conservative of the above methods.

The NLO corrections are as large as $70$\% relative to the LO
result.  The size of this correction stems from the unnaturally large residue of the
deuteron pole, $Z_t-1\approx0.7$, in Z-parameterization. However, once this effect is taken into
account, convergence at \NtwoLO and higher is actually improved~\cite{Phillips:1999hh,Beane:2000fx,Griesshammer:2004pe}. 
Different inputs to determine the PC 3NI affect only the
$\twoS\to X$ partial-wave amplitudes. At NLO, this can lead to a change of up
to $15\%$ in the amplitudes $\calM_\text{R}[\twoS\to Y;k]$, in line with the
error-estimate criterion (iii).

In the functions $c[(X-Y)](\Lambda)$ multiplying the PV couplings $\daR$ and $\deR$,
Fig.~\ref{fig:spinrotresults-nd}, contributions involving the \twoS-wave are
doubly suppressed relative to those which contain the \fourS-wave. Not only are their relative weights in
Eq.~\eqref{eq:poshelampparameterised} small, they are also small in
absolute size, as seen from the fact that the \twoS-wave scattering length itself is a factor of
$\approx10$ smaller than the \fourS one. Therefore, applying criterion (iii) leads to very small
variations. Varying the cutoff
$\Lambda\in[200;5000]\;\MeV$, criterion (ii), produces a range at NLO of
$c[(^3 \! S_1-^1 \! P_1)](\Lambda)=[7.8\dots8.1]\;\text{rad}\;\MeV^{-\half}$  and
$c[(^3 \! S_1-^3 \! P_1)](\Lambda)=-[16.5\dots17.4]\;\text{rad}\;\MeV^{-\half}$.  For error-estimate
criterion (i), i.e.~$Q^2\approx0.1$ of the total result, corrections are
$\pm0.8\;\text{rad}\;\MeV^{-\half}$ and $\pm1.7\;\text{rad}\;\MeV^{-\half}$,
significantly larger than the estimate from varying $\Lambda$. We therefore
adopt the range from criterion (i) as a conservative estimate of the theoretical
uncertainties in these coefficients.  Overall, the amplitude
$\calM_\text{R}[\fourS\to\wave{4}{P}{\frac{3}{2}}]$ dominates $c[(^3 \! S_1-^3 \! P_1)](\Lambda)$,
providing more than  $80$\% of its total value.

Due to symmetries, the function $c[\calT](\Lambda)$ does not receive any contributions from the $\fourS\to\fourPthree$ channel, and the amplitude $\calM_\text{R}[\fourS\to\wave{2}{P}{\frac{3}{2}}]$ is very small. The $\twoS\to\wave{4}{P}{\frac{1}{2}}$ amplitude thus dominates $c[\calT](\Lambda)$. Since this amplitude does not depend
significantly on the input used to determine the PC 3NI, criterion (iii)
significantly underestimates the theoretical uncertainty of the NLO
calculation. Cutoff-variation, criterion (ii), is significant in this channel, mapping out the
range $c[\calT](\Lambda)=[1.8\dots2.8]\;\text{rad}\;\MeV^{-\half}$,  while criterion (i)
provides an error-estimate of only $\pm 0.3\;\text{rad}\;\MeV^{-\half}$. We therefore adopt
the variation from the cutoff changes as a conservative estimate of the theoretical
uncertainties.  While this error is $\approx20\%$, the magnitude of $c[\calT](\Lambda)$ is at most a third of the other two $c[(X-Y)](\Lambda)$. If there are no accidental cancellations or significant differences in the magnitudes of the PV couplings themselves, the overall contribution of $c[\calT](\Lambda)$ to the spin rotation angle is expected to be small.

\section{Rotation angle estimates and comparisons}\label{Sec:Comp}
\setcounter{equation}{0}

Our results for $np$ and $nd$ spin rotation are given in terms of the LECs $g^{(X-Y)}$ of the
Lagrangian~(\ref{eq:PVLag}). As discussed in the Introduction, these couplings  are not predicted by the EFT, but can be estimated on dimensional
grounds. Since on the microscopic level the dominant PV process stems from a
single weak gauge boson exchange, the PV $NN$ interactions are suppressed
relative to the PC ones by a factor $\sim\frac{\gamma_t^2}{M_W^2}$, where
$\gamma_t$ is taken as a typical low-momentum scale. As in the PC
interactions between two nucleons and the auxiliary dibaryon field in
Eq.~\eqref{eq:y}, an overall factor of $\sqrt{\frac{4\pi}{M}}$ should be
included. In order to match the dimensionality $[\MeV^{-\frac{3}{2}}]$ of the
couplings $g^{(X-Y)}$, an additional inverse mass dimension enters.  Since
the PV couplings used here are renormalization-group invariant and encode
short-distance physics, the mass scale can only be set by the breakdown scale
$\bar{\Lambda}\approx m_\pi \approx140\;\MeV$ of \EFTNoPion. Overall, therefore, we expect the magnitude of the PV couplings to be of
order
\begin{equation}
\label{eq:couplingsestimate}
\left|g^{(X-Y)}\right|\sim\sqrt{\frac{4\pi}{M}}\;\frac{1}{\bar{\Lambda}}
\left(\frac{\gamma_t}{M_W}\right)^2\approx 10^{-10}\;\MeV^{-\frac{3}{2}}\;.
\end{equation}
An estimate which is compatible with this number may be obtained by appealing
to {Ref.}~\cite{Phillips:2008hn}, where a value for
the combination of couplings involved in PV $\vec p p$
scattering at 13.6 MeV \cite{318678} was extracted.
(Asymmetries measured at higher energies are not within the realm of
validity of \EFTNoPion.)
Using this value as an estimate of all LECs,
\begin{equation}
\label{eq:CouplingsExpEstimate}
\left|g^{(X-Y)}\right|\approx\ 2
\times 10^{-11}\;\MeV^{-\frac{3}{2}}
\end{equation}
in our conventions. Note that these are only rough estimates.

With these values for the magnitude of the PV couplings and a target density of
$\rho\approx10^{23}\;\text{cm}^{-3}$, the magnitude of the spin-rotation
signal in hydrogen of Eq.~\eqref{eq:npspinrotresultNLO} is
\begin{equation}
  \label{eq:npspinrotestimate}
  \left|\frac{\dd \phi_\text{PV}^{np}}{\dd l}\right|\approx \left[10^{-7} \cdots 10^{-6}\right]\;
  \frac{\text{rad}}{\text{m}} \;\;.
\end{equation}  
Analogously, the magnitude of the spin-rotation signal in deuterium of Eq.~\eqref{eq:ndspinrotresult} is roughly
\begin{equation}
  \label{eq:ndspinrotestimate}
  \left|\frac{\dd \phi_\text{PV}^{nd}}{\dd l}\right|\approx
  \left[10^{-7} \cdots 10^{-6}\right]\;\frac{\text{rad}}{\text{m}} \;\;.
\end{equation}  
We stress again that -- without reliable values of the PV parameters -- these results
are dimensional, order-of-magnitude estimates and may well be off by factors
of $10$ or more.

Another order-of-magnitude estimate can be obtained by using the parameter set of Ref.~\cite{arXiv:1110.1039}
extracted from the DDH ``best estimates''; using these values (Eq.~(31) of Ref.~\cite{arXiv:1110.1039}) and adjusting for different sign conventions yields spin rotation values of
\begin{equation}
 \frac{\dd \phi_\text{PV}^{np}}{\dd l}\approx 3 \times 10^{-7}\;
  \frac{\text{rad}}{\text{m}} \;\;.
\end{equation}
for $np$ spin rotation and 
\begin{equation}
 \frac{\dd \phi_\text{PV}^{nd}}{\dd l}\approx 5 \times 10^{-7}\;
  \frac{\text{rad}}{\text{m}} \;\;.
\end{equation}
in the $nd$ case. Note that there is a large spread in the DDH ``reasonable ranges'' surrounding these DDH ``best estimates.''

We now compare our estimated result with the results from other calculations. Without measured PV parameters, none of these can be considered as more than order-of-magnitude estimates. Hence we do not normalize with respect to less than order-of-magnitude changes in choice of target density, for example.

Our estimate for the $np$ spin rotation angle agrees well with earlier results obtained in the DDH and hybrid formalisms \cite{Avishai:1985mu,Schiavilla:2004wn,Zhu:2004vw,Liu:2006dm}, which found rotation angles between $5.15 \times 10^{-7}\frac{\text{rad}}{\text{m}}$ and $1.36 \times 10^{-6}\frac{\text{rad}}{\text{m}}$, depending on model choices made for the couplings and the strong interaction potentials. 

The rotation angle in neutron deuteron spin rotation was determined in
Ref.~\cite{Schiavilla:2008ic} using the DDH model and the Argonne
$v_{18}$ interaction, supplemented with the Urbana-IX three-nucleon
potential. The result is
\begin{equation}
\frac{\dd \phi_\text{PV}^{nd}}{\dd l} = 9.32 \times 10^{-7}
\frac{\text{rad}}{\text{m}}\;\;, 
\end{equation}
while a subsequent calculation also employing the A$v_{18}$+UIX
potentials found
\cite{Song:2010sz},
\begin{equation}
\label{eq:SongDDH}
\frac{\dd \phi_\text{PV}^{nd}}{\dd l} = 7.68 \times 10^{-7}
\frac{\text{rad}}{\text{m}}\;\;. 
\end{equation}
This agrees with our findings that the coefficients of the PV couplings in
$np$ and $nd$ rotations are of the same order of magnitude, and also agrees
with our rough estimate of Eq.~\eqref{eq:ndspinrotestimate}.
Note that result \eqref{eq:SongDDH} is just one of several $nd$ spin rotation
predictions provided by Ref.~\cite{Song:2010sz}, who consider two
other parameter set estimations in addition to a set with the ``best values'' from the wide ranges provided by DDH.  These two parameter sets, collected
by Bowman \cite{Bowman}, yield values of
  $-6.82 \times 10^{-7}$ and $-8.91 \times 10^{-7} \, \frac{\text{rad}}{\text{m}}$.

In addition, Refs.~\cite{Schiavilla:2008ic} and \cite{Song:2010sz} performed a
hybrid calculation in which the phenomenological wave functions of the
PC sector are combined with a PV potential derived from
\EFTNoPion.
In this approach the consistency of treating interactions, wave functions, and currents within a unified framework is lost. In particular, different degrees of freedom appear in the PC and PV interactions.
In both references, results for two different values of the
regularization parameter $\mu$ used in the PV \EFTNoPion potential are given, while in the PC sector a particular pionful potential with a fixed parameter set is used. 
For the choices $\mu=138\,\MeV$ and
$\mu=1\,\GeV$, the results for the coefficients multiplying the PV parameters
differ by up to two orders of magnitude. 
While this regularization dependence can in principle be removed by running the PV couplings to absorb cutoff dependence, these large differences are an indication of the resolution mismatch between \EFTNoPion and the phenomenological potentials. Also note that different regularization schemes and degrees of freedom are employed for PC and PV interactions.
Since the
renormalization-scale dependence of the PV and PC couplings in the hybrid formalism is unknown, one cannot determine whether the calculation is indeed
cutoff-independent, and varying the cutoff cannot be used to assess residual
theoretical uncertainties. On the other hand, as shown above, our results are
independent of any choices related to regularization and renormalization
within the errors estimated in the EFT approach.

\section{Conclusion and Outlook}\label{Sec:Outlook}
\setcounter{equation}{0}

At NLO in parity-violating \EFTNoPion, two- and three-body low-energy
PV observables depend upon five unknown LECs, the $g^{(X-Y)}$ of
Eq.~\eqref{eq:PVLag}.  PV \EFTNoPion is based on the symmetries
of QCD and the weak interaction, and the power counting of the theory.
In order to verify that QCD is appropriately encoded
into this EFT, it must be demonstrated that
the LECs extracted from independent observables are consistent and are
of ``natural'' size -- ${\cal O}(1)$ in dimensionless units.  To
overconstrain the system requires more than five
model-independent calculations involving linearly independent
combinations of the LECs, and the availability of the corresponding
measurements.

This paper provides two of these calculations: 
Using \EFTNoPion consistently in both PV and PC sectors of the nucleon
interactions, we obtain model-independent
results for neutron spin rotation in hydrogen and
deuterium targets. 
At NLO they are given by
\begin{align*}
\frac{1}{\rho}\;\frac{\dd \phi_\text{PV}^{np}}{\dd l} =  &\left( [4.5 \pm 0.5]
  \, (2\deR+\daR) \right.\\
  &\left. - [18.5 \pm 1.9] \, (\dbR - 2 \ddR) \right) {\rm rad \ MeV^{-\half}}
\end{align*}
for $np$, and
\begin{align*}
\frac{1}{\rho} \frac{\dd \phi_\text{PV}^{nd}}{\dd l}
  =  &\left( [8.0\pm0.8]\;\daR\;+\;[17.0\pm1.7]\;\deR \right.\\
  &\left. +[2.3\pm0.5]\;(3\dbR -2\dcR) \right){\rm rad \ MeV^{-\half}}
\end{align*}
for $nd$.
They yield two independent constraints on
the five LECs. Absent any cancellation between different PV
parameters we estimate that the rotation angles for both hydrogen and deuterium
targets are of the same size,
\begin{equation}
  \left|\frac{\dd \phi_\text{PV}}{\dd l}\right|\approx
  \left[10^{-7} \cdots 10^{-6}\right]\;\frac{\text{rad}}{\text{m}} \;\;.
\end{equation} 
There is no indication that the spin rotation observable is
enhanced for a deuteron target, 
in agreement with Refs.~\cite{Schiavilla:2008ic,Song:2010sz}.
Numerical stability analyses verify the theoretical findings of
Ref.~\cite{Griesshammer:2010nd} that no parity-violating 3NI is necessary at
leading or next-to-leading order.

The two calculations presented here join three others published in the same framework; the longitudinal asymmetry in
$\vec{p} p$ scattering \cite{Phillips:2008hn} and two PV observables 
from the 
$np \leftrightarrow  d\gamma$ system \cite{Schindler:2009wd}. 
Note that other two-nucleon PV calculations using pionless EFT
can be found in Refs.~\cite{Zhu:2004vw,Savage:2000iv,Shin:2009hi}. To go beyond consistency to the realm of potential prediction, 
we will next consider PV
observables in $Nd$ scattering,
  $nd\longleftrightarrow {}^3\mathrm{H}\gamma$ and $pd\longleftrightarrow
  {}^3\mathrm{He}\gamma$ before considering heavier systems.

\section*{Acknowledgments}

While this paper was under review, J. Vanasse posted Ref.~\cite{arXiv:1110.1039}
with a result on spin rotation off deuterium. We thank him for discussions clarifying some differences between his and our results that led to the correction of an error in our calculations.
We thank D.~R.~Phillips for important suggestions and encouragement; M.~Snow
for extensive discussions on possible future PV experiments and his detailed
write-up of spin-rotation definitions; as well as V.~Gudkov, A.~Micherdzinska, A.~K.~Opper, 
M.~Paris, and Y.~-H.~Song for insightful discussions and helpful suggestions. 
We are particularly indebted to the organizers and participants of
the Department of Energy's Institute for Nuclear Theory (INT) program
10-01: ``Simulations and Symmetries'' at the University of Washington,
and for the Department of Energy's financial support of our visit.  HWG is grateful for the kind hospitality of the Nuclear Experiment
group of the Institut Laue-Langevin (Grenoble, France).  MRS thanks the
Lattice and Effective Field Theory group at Duke University for their
hospitality. This work was carried out in part under National Science
Foundation \textsc{Career} award PHY-0645498 (HWG, MRS), and US-Department of
Energy grants DE-FG02-95ER-40907 (HWG, MRS) and DE-FG02-05ER41368 (RPS). We also
acknowledge support by University Facilitating Funds of the George Washington
University (HWG), and by the Center for Nuclear Studies of the George
Washington University (HWG, MRS).


\appendix

\section{Partial-Wave Projectors}
\label{Sec:Projectors}
\setcounter{equation}{0}


In this Appendix we construct the projectors used to extract the desired
  partial wave state(s) from a given $N d_{s/t}$ state.  
The partial wave
  state is labelled $[{}^{2S+1}L_J,I]$, where $S$ is the total spin, $L$ the
  orbital angular momentum, $J$ the total angular momentum, and $I$ the
  isospin. The projectors given here do not exhaust what is needed to do a
  partial wave decomposition for higher order or inelastic calculations; that
  would require constructing projectors acting on $NNN$ states as well. We employ the cluster-configuration basis introduced in Ref.~\cite{Griesshammer:2004pe}, which also presents the S-wave projectors.

The
  $N^{a \alpha}$ field has two free indices: the SU(2) isospin index $a$ and
  the SU(2) spin index $\alpha$.  The dibaryon field $d^i_t$ is an isosinglet
  and spin-triplet (\threeS) with a free vector index $i$, while $d^A_s$ is an
  isovector and spin-singlet (\oneS) with a free isovector index $A$.  To
  obtain a total $S=\half$ from an $Nd_{s/t}$ combination (indices
  suppressed), we can consider $Nd_s$, which is already purely $S=\half$ from
  the SU(2) decomposition $\mathbf{\half}\otimes\mathbf{0}=\mathbf{\half}$.
  An $S=\half$ term is also available from $Nd_t$ via
  $\mathbf{\half}\otimes\mathbf{1}=\mathbf{\frac{3}{2}}\oplus\mathbf{\half}$.
  The $S=\half$ piece is isolated by contracting with a Pauli matrix in spin
  space: $\frac{1}{\sqrt{3}}(\sigma_i)^\mu_\alpha N^{a \alpha} d^i_t $.
  Summation of repeated indices is implied.  The normalization chosen will be
  explained at the end of this Section.

An $S= \frac{3}{2}$ term can only be obtained from $N d_{t}$.   It has four degrees of freedom in spin space: one free vector index (the $i$ in $d^i_t$) with one free SU(2) spin index (the 
$\alpha$ in $N^{a \alpha}$)
 gives six degrees of freedom, but the two that contribute to
 $S=\half$ are removed by additional constraints on the projector.
 The most general form is given by
\begin{equation}
 Q^i{}_j=a \delta^i_j+b \sigma^i\sigma_j.
\end{equation}
Requiring that the $S=\half$ and $S=\frac{3}{2}$ projectors are orthonormal
\begin{equation} 
\sigma_i\,Q^i{}_j=0=Q^i{}_j\,\sigma^j\,, \quad Q^i{}_k\,Q^k{}_j=Q^i{}_j\,, 
\end{equation}
yields
\begin{equation}
\label{eq:Qij}
  Q^i{}_j=\delta^i_j-\frac{1}{3}\sigma^i\sigma_j=
  \frac{1}{3}\left[2\delta^i_j-\ii\,\epsilon^{i}{}_{jk}\,\sigma^k\right],
\end{equation}
where the second expression uses
$\sigma^i\sigma_j=\delta^i_j+\ii\,\epsilon^i{}_{jk}\sigma^k$. 
Orthogonality to the $S=\half$ projector results in the required constraints
that reduce the number of degrees of freedom to four. The projector onto
$S=\frac{3}{2}$ satisfies
\begin{equation}
 \left(Q^i{}_j\right)^\dagger=Q^j{}_i\,.
\end{equation}

For \wave{}{P}{}-waves, the projector is obtained by combining $L=1$ powers of
the unit vector $\ev$ in the direction of the $d_{s/t}$ center-of-mass
momentum with the auxiliary field $d_{s/t}$ and the nucleon. In the case of the spin-1
field $d_t$, this leads to total angular momentum
$\mathbf{1}\otimes\mathbf{1}\otimes\mathbf{\half}=
\mathbf{\frac{5}{2}}\oplus\mathbf{\frac{3}{2}}\oplus\mathbf{\half}$. The
projector onto $J=\frac{5}{2}$ is:
\begin{equation}
Q^{ij}{}_{kl}:=
\frac{9}{10}\,\left[\left(\delta^i_k\delta^j_l+\delta^i_l\delta^j_k
        -2\delta^{ij}\delta_{kl}\right)
      -\frac{\ii}{3}\left(
         \epsilon^{i}{}_{km}\delta^j_l+\epsilon^{j}{}_{km}\delta^i_l
+\epsilon^{i}{}_{lm}\delta^j_k+\epsilon^{j}{}_{lm}\delta^i_k\right)\sigma^m
    \right]\;,
\end{equation}
which is symmetric and traceless in $(ij)$ and $(kl)$ separately,
\begin{equation}
  \delta_{ij}Q^{ij}{}_{kl}=\delta^{kl}Q^{ij}{}_{kl}=0\;\;,\;\;
  Q^{ij}{}_{kl}=Q^{ji}{}_{kl}\;\;,\;\;Q^{ij}{}_{kl}=Q^{ij}{}_{lk}\;\;,
\end{equation}
\begin{equation}
  \left(Q^{ij}{}_{kl}\right)^\dagger=Q^{kl}{}_{ij}\;\;,
\end{equation}
and orthonormal to the $S=\half$ and $S=\frac{3}{2}$ projectors above:
\begin{equation}
  \sigma_i\,Q^{ij}{}_{kl}=Q^{ij}{}_{kl}\,\sigma^k=0\;\;,\;\;
  Q^m{}_i\,Q^{ij}{}_{kl}=Q^{ij}{}_{kl}\,Q^l{}_m=0\;\;,\;\;
  Q^{ij}{}_{mn}\,Q^{mn}{}_{kl}=Q^{ij}{}_{kl}.
\end{equation}
Projectors onto definite states are obtained by contracting $\sigma_j$ and
$Q^i{}_j$ with $d_{s/t}N$ to extract the desired spin-state and multiplying by
$L$ powers of momentum. Any remaining free indices are then contracted with
the appropriate projector onto the desired total angular momentum, which is
again given by $\sigma_j$, $Q^i{}_j$, or $Q^{ij}{}_{kl}$, and with
Kronecker-Deltas or $\epsilon^{ijk}$ as needed.  For example,
$(\sigma_ke^k)(\sigma_ld_t^lN)$ has the quantum numbers $S=\half,J=\half,L=1$,
i.e. \wave{2}{P}{\half}. Or, as a more complex example:
\wave{4}{P}{\frac{3}{2}} mandates first coupling $d_t^lN$ to $S=\frac{3}{2}$
using $Q^m{}_l$, resulting in $Q^m{}_ld_t^lN$, and then multiplying by $e_k$.
The two free vector indices have to be contracted with another projector
$Q^i{}_n$, multiplied from the left, such that a $J=\frac{3}{2}$ state
results. This can only be achieved by multiplying with $\epsilon^{nk}{}_m$.
The projector is thus proportional to
$Q^i{}_n\,\epsilon^{nk}{}_m\,e_k\,Q^m{}_l d_t^lN$.

The same construction principle holds for isospin projections. States with
$I=\half$ can be generated by combining either $d_t$ ($I=0$) or $d_s$ ($I=1$)
with the $I=\half$ field $N$, since as before
$\mathbf{0}\otimes\mathbf{\half}=\mathbf{\half}$ and
$\mathbf{1}\otimes\mathbf{\half}=\mathbf{\frac{3}{2}}\oplus\mathbf{\half}$. To
obtain an $S=\half,I=\half$ state, first contract all spin-vector indices of
$d_t^l N^{a\alpha}$ using $\sigma_l$, or all isospin-vector indices of $d_s^A
N^{a\alpha}$ using $\tau_A$, leading to two equivalent terms
\begin{equation}
  \left[\left(\sigma_i\right)^{\alpha}{}_{\beta}\,\delta^a_b\, d_t^i+
    \left(\tau_A\right)^{a}{}_{b}\,\delta^\alpha_\beta\,d_s^A 
  \right]N^{b\beta}\equiv
  \left[\sigma_i\, d_t^i+\tau_A\,d_s^A \right]N.
\end{equation}
Here, all spin and isospin indices are listed explicitly in the first
expression, while obvious index contractions are suppressed in the second. As
two cluster-configurations exist, namely $d_tN$ and $d_sN$, it is convenient
to follow Ref.~\cite[App.~A.1]{Griesshammer:2005ga} in decomposing all
operators as 
\begin{equation}
  \calO = N^\dagger_{b\beta} \;\left(d^\dagger_{t,j},\;d^\dagger_{s,B}\right)
  \begin{pmatrix}\calO(Nd_t\to Nd_t)^j_i&\calO(Nd_s\to Nd_t)^j_A\\
    \calO(Nd_t\to Nd_s)^B_i&\calO(Nd_s\to Nd_s)^B_A
  \end{pmatrix}^{b\beta}_{\hq\hq a\alpha} \;\binom{d_t^i}{d_s^A}\;N^{a\alpha}.
\end{equation}
Each operator is represented in the cluster-configuration basis by a 2x2-matrix
which carries spin and isospin indices, and all operators act in the direct
tensor product space
$\textbf{spin}\otimes\textbf{isospin}\otimes\textbf{cluster}$.  The operator in the cluster-configuration basis that projects onto the $S=\half,
I=\half$ component is
\begin{equation}
  (\calP_{d,iA})^{m\mu}{}_{a\alpha}=\frac{1}{\sqrt{3}}\;\begin{pmatrix}
    \sigma_i&0\\0&\tau_A\end{pmatrix}^{m\mu}_{\hq\hq a\alpha}\;\;.
\end{equation}  
On the other hand, only $d_tN$ contains an $S=\frac{3}{2},I=\half$ component,
so that it is useful to define 
\begin{equation}
  (\calP_{q,j}^i)^{m\mu}{}_{a\alpha}=\begin{pmatrix}
    Q^i{}_j&
    0\\0&0\end{pmatrix}^{m\mu}_{\hq\hq a\alpha}= Q^i{}_j\,\calP_q \;\;
  \mbox{ with } \calP_q:=\begin{pmatrix}1& 0\\0&0\end{pmatrix}\;\;.
\end{equation}
Here, $\calP_q$ is the matrix projecting onto the only physical component of
the $S=\frac{3}{2}$ cluster-configuration matrix. The 
following relations hold:
\begin{equation}
\begin{split}
  \calP_d^{iA} &=\left(\calP_{d,iA}\right)^\dagger=
  \mbox{``}\calP_{d,iA}\mbox{''}\;, \quad 
 \calP_{d,iA}\calP_{d}^{iA} =\mathbbm{1}\;, \\
  \left(\calP_{q,i}^j\right)^\dagger &=\calP_{q,j}^i\;, \quad 
 \calP_{q,j}^k\,\calP_{q,i}^j =\calP_{q,i}^k \calP_{d,iA}\,\calP_{q,j}^i=0\;.
\end{split}
\end{equation}
The unit matrix in spin and isospin space is
$(\mathbbm{1})^{b\beta}{}_{a\alpha}=\delta^\beta_\alpha\,\delta^b_a$.


Below are the projectors as sources of the fields $N$ and $d_{s/t}$ with the
desired quantum numbers. 
First, in analogy with the auxiliary two-nucleon fields $d_{s/t}$, we introduce 
three-nucleon interpolating fields with the quantum numbers of the $Nd_{s/t}$
state. In symbolic notation, they are the cluster-configuration vectors
representing sources  $\text{T}[{}^{2S+1}{L}_{J},I]^\dagger$ with total spin $S$,
orbital angular momentum $L$, total angular momentum $J$,
isospin $I$, and appropriate vector and spinor indices $\{M\}$.
The projection onto a definite partial wave in cluster-configuration space is
then
\begin{equation}
\label{eq:projectonsource}
  \left(\text{T}[{}^{2S+1}{L}_{J},I]^\dagger\right)_{\{M\}}\;
  \calP^{\{M\}}{}_{lA}[{}^{2S+1}{L}_{J},I]\;\binom{d_t^l}{d_s^A}\;N\;\;.
\end{equation}
The complete set of S- or P-wave projectors using auxiliary fields as source
states is finally:
\begin{equation}
\begin{array}{lll}
  \calP[\wave{2}{S}{\frac{1}{2}},I=\half]_{lA}&=&\dis\calP_{d,lA}\ ,\\[1ex]
  \calP[\wave{2}{P}{\frac{1}{2}},I=\half]_{lA}&=&
    \dis\left(\vec{\sigma}\cdot{\ev}\right)\calP_{d,lA}\ ,\\[1ex]
  \calP[\wave{2}{P}{\frac{3}{2}},I=\half]^i{}_{lA}&=&
    \dis\sqrt{3}\;Q^i{}_k\,e^k\,\calP_{d,lA}\ ,\\[1ex]
  \calP[\wave{4}{S}{\frac{3}{2}},I=\half]^i{}_{l}&=&\dis\calP_{q,l}^i\equiv
  Q^i{}_l\calP_q\ ,\\[1ex]
  \calP[\wave{4}{P}{\frac{1}{2}},I=\half]{}_{l}&=&
    \dis\sqrt{\frac{3}{2}}\,e_k\,Q^k{}_l\,\calP_q   \ , \\[2ex]
  \calP[\wave{4}{P}{\frac{3}{2}},I=\half]^i{}_{l}&=&
    \dis\frac{3\ii}{\sqrt{5}}\,Q^i{}_n\,\epsilon^{nk}{}_m\,e_k
    \,Q^m{}_l\,\calP_q=
    \dis\frac{1}{3\sqrt{5}}\left[\sigma^i\,\delta^k_l+\sigma_l\,\delta^{ik}
      -4\sigma^k\,\delta^i_l-5\ii\,\epsilon^{ik}{}_{l}\right]e_k\,
    \calP_q \ ,\\[2ex]
    \calP[\wave{4}{P}{\frac{5}{2}},I=\half]^{ij}{}_{l}&=&
    \dis Q^{ij}{}_{kl} e^k\,\calP_q\ .
\end{array}
\end{equation}
The spin-quartet projectors carry no isovector index; the normalization is
discussed below.

The two expressions for $\calP[\wave{4}{P}{\frac{3}{2}}]$ are equivalent since
$Q^m{}_{l}$ contains two Pauli matrices and is multiplied with one more from
the left, and products of Pauli matrices can be reduced to a sum of terms
containing at most one Pauli matrix. The first form contains the spin-quartet
projector $\calP_{q,l}^m\equiv Q^m{}_l\calP_q$ explicitly and hence is
manifestly orthogonal to the spin-doublet projectors. The second form contains
the minimal number of linearly independent structures, derived by building the
most general matrix $A^i{}_{kl}$ out of $\sigma^i$, $\delta^i_j$ and
$\epsilon^{ijk}$.

The orthonormalization condition was imposed as follows: Projectors to
different partial waves are orthogonal after contraction over the
auxiliary-field variable $(lA)$ and integration over the solid angle element
$\dd\Omega_e$ of the auxiliary field cm-momentum direction $\ev$:
\begin{equation}
  \frac{1}{4\pi}\int\deint{}{\Omega_e}
  \calP[{}^{2S+1}L_J,I]^{\{M\}}{}_{lA}\;
  \left(\calP[{}^{2S^\prime+1}L^\prime_{J^\prime},I^\prime]^\dagger
  \right)_{\{N\}}^{lA}=0
  \;\;\forall\;\; L\not=L^\prime \vee S\not=S^\prime\vee J\not=J^\prime
  \vee I\not=I^\prime.
\end{equation}
When the states are identical, integration
and contraction must yield the identity element in the space of total angular
momentum states. In the cluster-configuration basis notation:
\begin{align}
  \text{for $J=\frac{1}{2}$:}& \quad \frac{1}{4\pi}\int\deint{}{\Omega_e}
  \calP[{}^{2S+1}L_\half,I]_{lA}\;\calP^\dagger[{}^{2S+1}L_\half,I]^{lA}=\mathbbm{1} \ ,\\ 
  \text{for $J=\frac{3}{2}$:}& \quad \frac{1}{4\pi}\int\deint{}{\Omega_e}
  \calP[{}^{2S+1}L_\frac{3}{2},I]^i{}_{lA}\;
  \calP^\dagger[{}^{2S+1}L_\frac{3}{2},I]^{lA}{}_j=\begin{cases}
    Q^i{}_j&\text{for } S=\half \ ,\\
    Q^i{}_j\,\calP_{q}&\text{for } S=\frac{3}{2}\ ,\end{cases}
  \\
  \text{for $J=\frac{5}{2}$:}& \quad \frac{1}{4\pi}\int\deint{}{\Omega_e}
  \calP[{}^{2S+1}L_\frac{5}{2},I]^{ij}{}_{mn}\;
  \calP^\dagger[{}^{2S+1}L_\frac{5}{2},I]^{mn}{}_{kl}= Q^{ij}{}_{kl}\,\calP_{q} \;.
\end{align}

\section{Assessing choices for the PC amplitudes of the class-II diagrams}
\label{app:insertions}
\setcounter{equation}{0}

\begin{figure}[!htbp]
  \begin{center}
    \includegraphics*[width=0.96\linewidth]{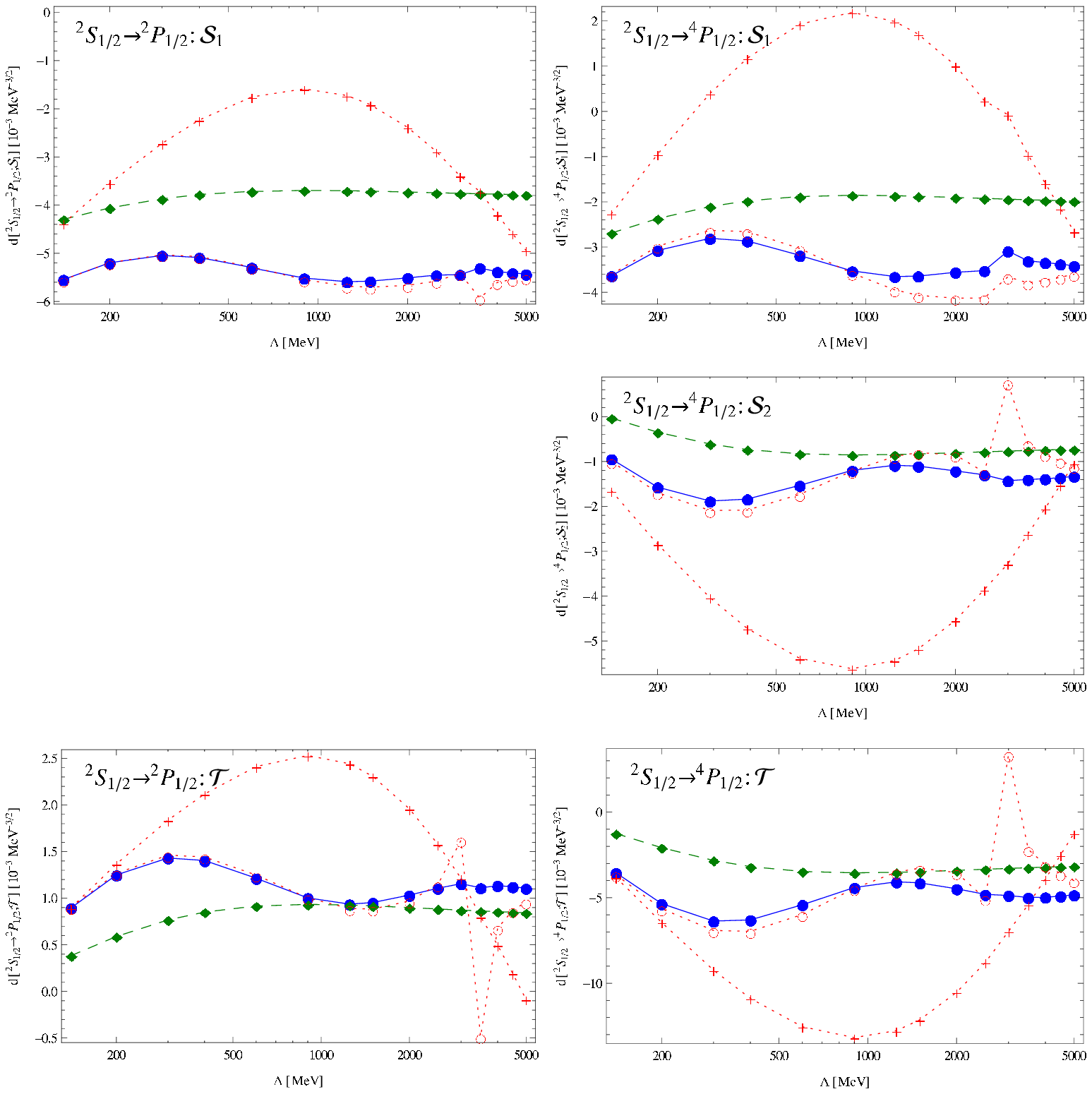}
     \caption{Cutoff dependence of the functions
      $d[\twoS\to Y;\text{coupling}]$, Eq.~\eqref{eq:ndPWdecomposed}, for different choices of the order
      of the PC amplitudes in the class-II contributions of
      Fig.~\ref{fig:PV3NstrictNLOsameloop} and
      Eqs.~(\ref{eq:1loopinsertionleft}/\ref{eq:1loopinsertionright}/\ref{eq:2loopinsertion}).
      \protect\gdiamond
      (dashed lines): LO result, 3NI $H_0$ from $nd$
      \twoS-scatt.~length; all other: NLO results with different treatments of
      PC amplitudes, with 3NI from \twoS-scatt.~length. \bdisc (solid
      lines): PC amplitudes \twoS as NLO, \wave{2,4}{P}{} as LO
      (Eq.~\eqref{eq:insertionchoice}); \rcirc (dotted lines): all as NLO; \rcross
      (dotted lines): all as LO.  Linear extrapolations used to guide the eye.
      Notice the different scales on the vertical axes.  Amplitudes which are
      identically zero are not displayed.  Columns (left to right): partial
      waves $\twoS\to\wave{2}{P}{\half}$, $\twoS\to\wave{4}{P}{\half}$; rows
      (top to bottom): coefficients of $\calS_1$, $\calS_2$, $\calT$.}
    \label{fig:partialwaveresults-nd-compareinsertions-2S}
  \end{center}
\end{figure}

\begin{figure}[!htbp]
  \begin{center}
    \includegraphics*[width=0.96\linewidth]{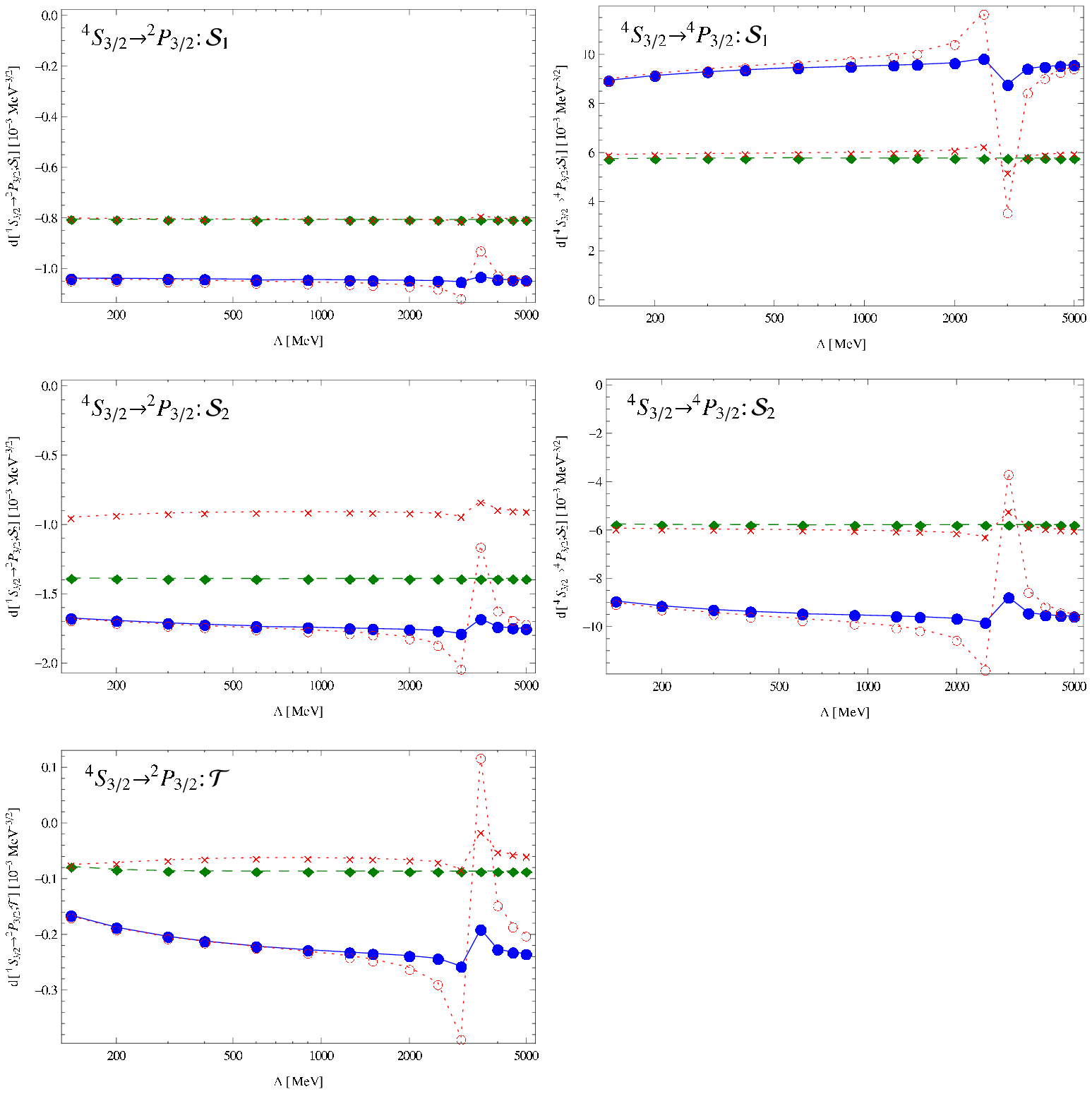}
    \caption{Cutoff dependence of the expansion coefficients
      $d[\fourS\to Y;\text{coupling}]$, Eq.~\eqref{eq:ndPWdecomposed}, for different choices of the order
      of the PC amplitudes in the class-II contributions of
      Fig.~\ref{fig:PV3NstrictNLOsameloop} and
      Eqs.~(\ref{eq:1loopinsertionleft}/\ref{eq:1loopinsertionright}/\ref{eq:2loopinsertion}).
      \protect\gdiamond
      (dashed lines): LO result; all other: NLO results
      with different treatments of PC amplitudes. \bdisc (solid lines): PC
      amplitudes \fourS as NLO, \wave{2,4}{P}{} as LO (Eq.~\eqref{eq:insertionchoice}); \rcirc (dotted lines): all as NLO; \rx      (dotted lines): all as LO. Linear extrapolations used to guide the eye.
      Notice the different scales on the vertical axes.  Amplitudes which are
      identically zero are not displayed.  Columns (left to right): partial
      waves $\fourS\to\wave{2}{P}{\frac{3}{2}}$, $\fourS\to\wave{4}{P}{\frac{3}{2}}$; rows
      (top to bottom): coefficients of $\calS_1$, $\calS_2$, $\calT$.}
    \label{fig:partialwaveresults-nd-compareinsertions-4S}
  \end{center}
\end{figure}

\begin{figure}[!htbp]
  \begin{center}
    \includegraphics*[width=0.5\linewidth]{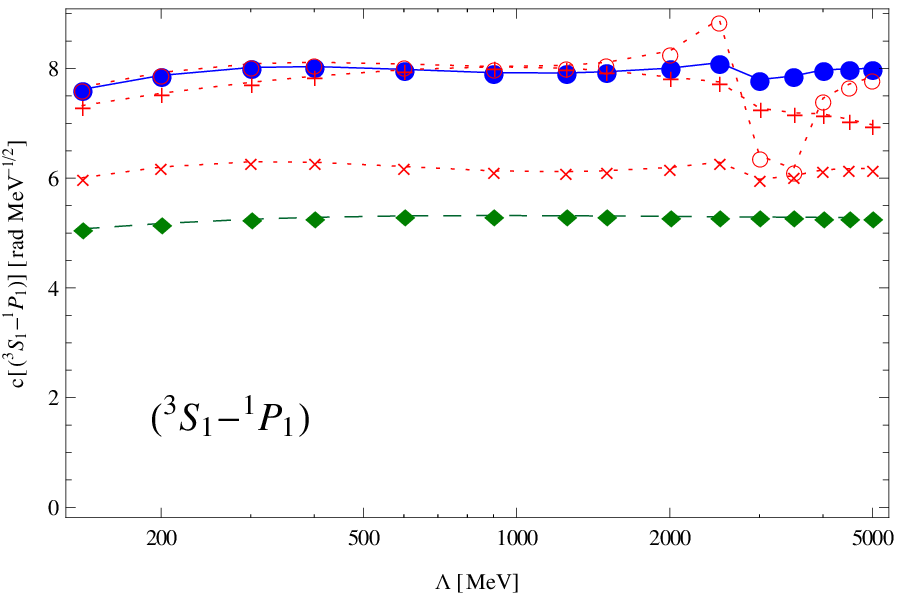}\\[1ex]
    \includegraphics*[width=0.5\linewidth]{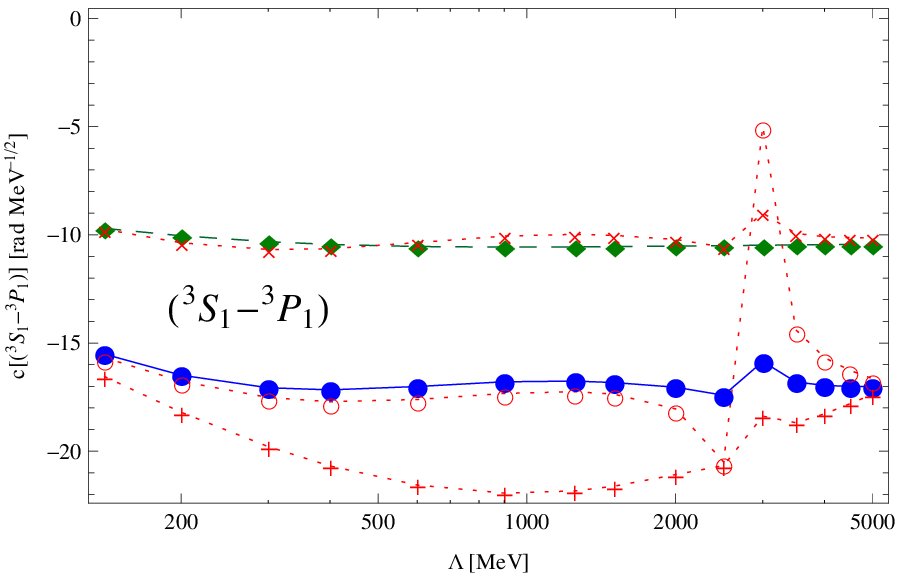}\\[1ex]
    \includegraphics*[width=0.5\linewidth]{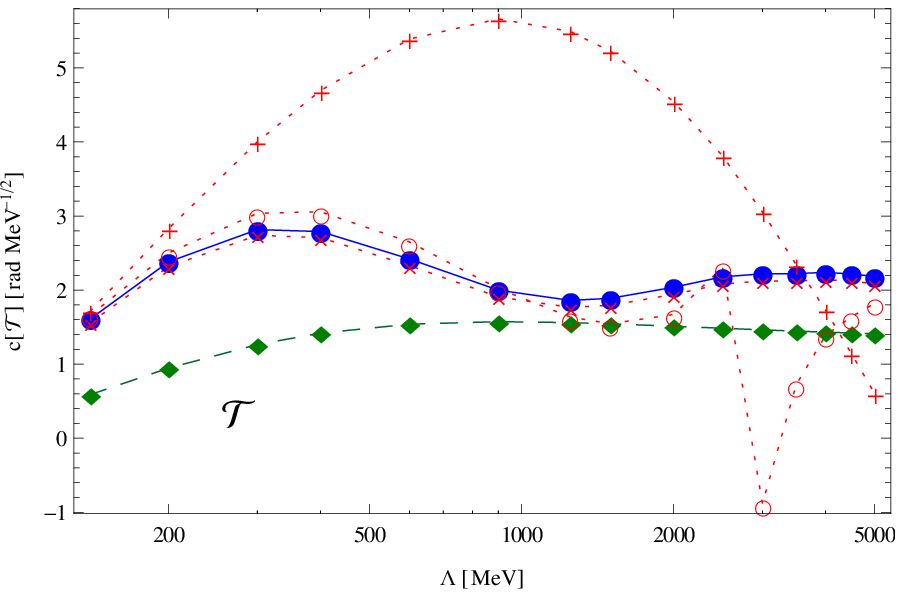}
    \caption{Cutoff dependence of the expansion coefficients
      $c[g^{(X-Y)}]$ of the neutron spin-rotation angle in deuterium,
      Eq.~\eqref{eq:ndspinrotdecomposed}, for different choices of the order of
      the PC amplitudes in the class-II contributions of
      Fig.~\ref{fig:PV3NstrictNLOsameloop} and
      Eqs.~(\ref{eq:1loopinsertionleft}/\ref{eq:1loopinsertionright}/\ref{eq:2loopinsertion}).
      Notation as in
      Figs.~\ref{fig:partialwaveresults-nd-compareinsertions-2S}
      and~\ref{fig:partialwaveresults-nd-compareinsertions-4S}.  Linear
      extrapolations to guide the eye; different scales on the vertical axes.}
    \label{fig:spinrotresults-nd-compareinsertions}
  \end{center}
\end{figure}

In this Appendix, we discuss in more detail the choices of the PC amplitudes $t[X;k,q]$ of Eq.~\eqref{eq:insertionchoice} in the
class-II contributions at NLO appearing in
Eqs.~(\ref{eq:1loopinsertionleft}/\ref{eq:1loopinsertionright}/\ref{eq:2loopinsertion})
and shown in Fig.~\ref{fig:PV3NstrictNLOsameloop}.
As discussed in the main text, the convolution kernel itself is already
NLO, so that in principle a LO PC amplitude suffices.
Figures~\ref{fig:partialwaveresults-nd-compareinsertions-2S},~\ref{fig:partialwaveresults-nd-compareinsertions-4S}
and~\ref{fig:spinrotresults-nd-compareinsertions} compare the
dependence of
the total amplitudes on the different treatments of $t[X;k,q]$,
benchmarked against the
leading-order result and the choice of
Eq.~\eqref{eq:insertionchoice} to include the \wave{2,4}{S}{}-wave
amplitudes as NLO and the \wave{2,4}{P}{}-wave amplitudes as LO, see
Eq.~\eqref{eq:insertionchoice}.

Consider first the inclusion of the PC \twoS-wave amplitude at LO
instead of NLO (\rcross in Fig.~\ref{fig:partialwaveresults-nd-compareinsertions-2S}), while treating all other amplitudes according to Eq.~\eqref{eq:insertionchoice}. This leads to
individual contributions diverging logarithmically and as $q^{0.23\dots}$, as
demonstrated on general grounds in Ref.~\cite{Griesshammer:2010nd} and
numerically confirmed here. 
The problem is evident in the $\Lambda$ dependence
of the functions $d$ and $c$ of Eqs.~\eqref{eq:ndPWdecomposed} and
\eqref{eq:ndspinrotdecomposed}, as shown in Figs.~\ref{fig:partialwaveresults-nd-compareinsertions-2S} and \ref{fig:spinrotresults-nd-compareinsertions}.
This choice must thus be discarded as not
properly renormalized.

Next, it is consistent to include the PC \fourS-wave amplitude at LO
instead of NLO (\rx in
Fig.~\ref{fig:partialwaveresults-nd-compareinsertions-4S}), while treating
all other amplitudes according to Eq.~\eqref{eq:insertionchoice}. This still
determines the PV amplitudes at NLO since the kernel of the convolution
already counts as NLO. With the exception of $d[\fourS\to \twoPthree,\calS_2]$, however,
the corresponding results are very close to those obtained in a LO
calculation of the PV amplitudes (\gdiamond). Using the PC \fourS-wave
amplitude at NLO (\bdisc and \rcirc) induces a considerable shift
of all functions $d[\fourS\to Y;\text{coupling}]$. As discussed in
Sec.~\ref{sec:ndNLO}, PC LO and NLO \fourS amplitudes predict values of the
\fourS scattering length that differ by about $25\%$, with the NLO result
very close to the experimental value. The PV amplitudes, and therefore
the $d[\fourS\to Y;\text{coupling}]$, are expected to be highly sensitive to the
\fourS scattering length. This is supported by the plots in
Fig.~\ref{fig:partialwaveresults-nd-compareinsertions-4S}. We therefore choose the NLO
amplitude for the PC \fourS wave. Higher-order corrections from the
\fourS-wave amplitude are expected to be small, since the NLO expression for
the scattering length is already in good agreement with experiment.

Finally, it is consistent to include the PC \wave{2,4}{P}{}-wave 
amplitudes at NLO (\rcirc in Figures) instead
of LO (\bdisc in Figures), while treating all other amplitudes according to Eq.~\eqref{eq:insertionchoice}. One
might speculate that this leads to significant changes since the zero-energy
effective-range parameter of the amplitudes, the scattering volume, is up to a
factor of 2 bigger at NLO than at LO; see Ref.~\cite{Griesshammer:2004pe}.
Instead, the amplitudes appear largely insensitive to the effective-range
parameters of the $\mathrm{P}$-waves.  However, the spurious cutoff dependence already seen when the P-wave amplitudes
are included at LO is widened and increased to a pole in the window
$\Lambda\in[2000\dots4000]\;\MeV$ in the total amplitudes.  
Its origin seems to be that the partially-resummed approach includes some
contributions beyond NLO which need to be renormalized by \wave{2,4}{P}{}-wave 3NIs which are not
present at NLO. Outside that window, the functions $c$ and $d$ agree with
the choice made in Eq.~\eqref{eq:insertionchoice} within the error-estimate of Sec.~\ref{sec:ndresult}.


\end{document}